\documentstyle[epsfig,subfigure,12pt,a4]{article}
\begin{document}
\newcommand{\vC}{Cherenkov}
\newcommand{\ry}{recovery}
\newcommand{\dt}{deployment}
\newcommand{\ol}{optical}
\newcommand{\me}{module}
\begin{titlepage}
\begin{center}
\epsfig{figure=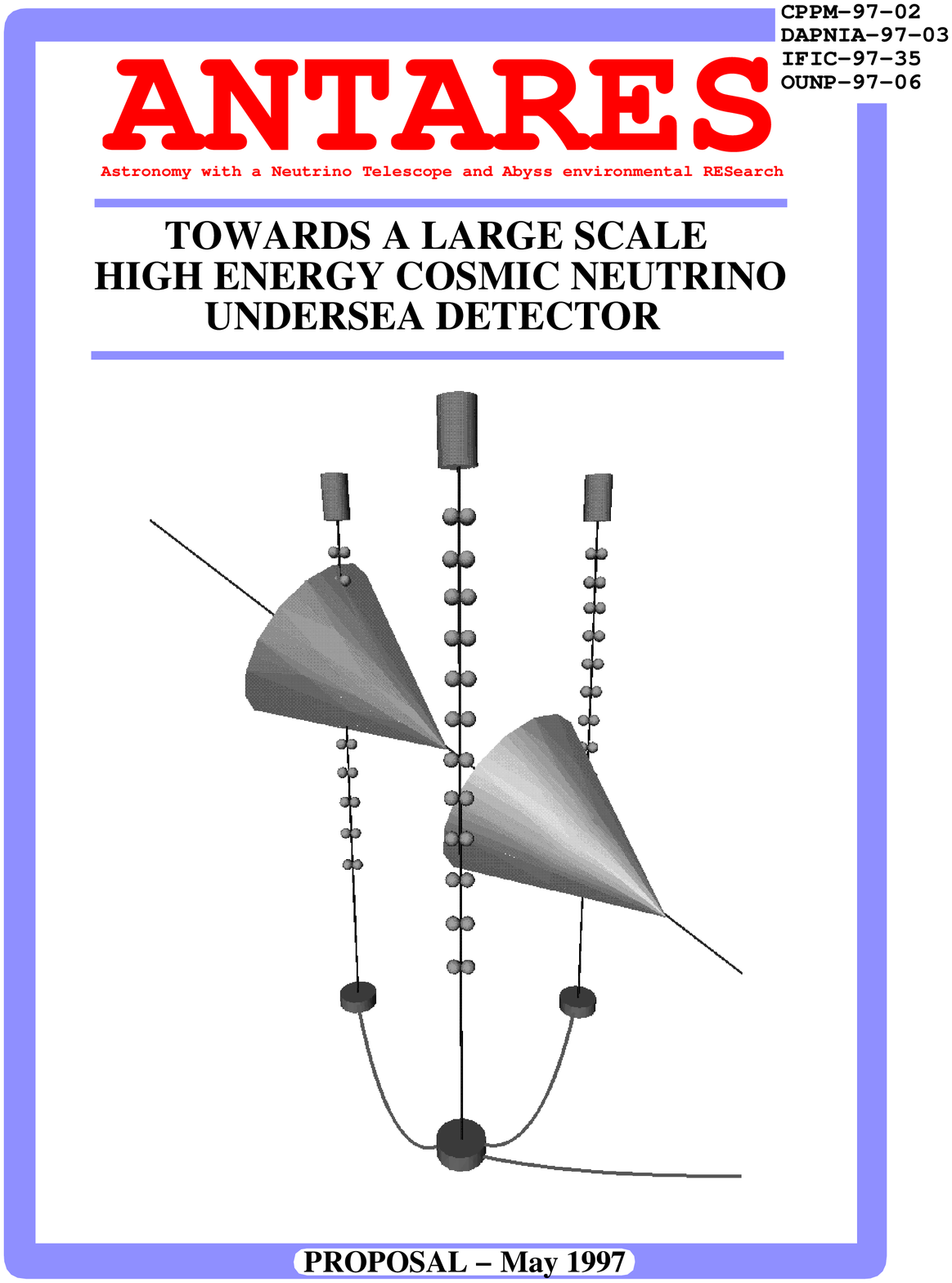,width=1.\linewidth}
\end{center}
\end{titlepage}
\newpage
\thispagestyle{empty}
\begin{flushright}
\fbox{\parbox{.8\linewidth}
{This document can be retreived as a postscript file 
via the web and internet on the CERN preprint server:\\
http://preprints.cern.ch/\\
or on the following servers\\
http://infodan.in2p3.fr/antares/docs.html\\
http://marcpl1.in2p3.fr/antares/docs.html\\
}}
\end{flushright}
\newpage
\thispagestyle{empty}
\begin{center}
{\LARGE \bf The ANTARES Collaboration}\\
\vglue 10mm
{\large \bf Centre d'Oc\'eanologie de Marseille}\\
{\normalsize \bf INSU - CNRS / Universit\'e de la M\'editerran\'ee}\\
\vglue 1mm
{\normalsize \bf F. Blanc, J.L. Fuda, L. Laubier, C. Millot.}
\\[5mm]         
{\large \bf Centre de Physique des Particules de Marseille}\\
{\normalsize \bf IN2P3 - CNRS / Universit\'e de la M\'editerran\'ee}\\
\vglue 1mm
{\normalsize \bf 
  C.~Arpesella, E.~Aslanides, J.J.~Aubert, S.~Basa, V.~Bertin, M.~Billault,
  P.E.~Blanc, A.~Calzas, C.~Carloganu,
  J.~Carr, J.J.~Destelle, F.~Hubaut, E.~Kajfasz, R.~Le~Gac,
  A.~Le~Van~Suu, L.~Martin, C.~Meessen, F.~Montanet,
  Ch.~Olivetto, P.~Payre, R.~Potheau, M.~Raymond, M.~Talby, E.~Vigeolas.
}
\\[5mm]
{\large \bf D\'epartement d'Astrophysique, de physique des Particules,\\
de physique Nucl\'eaire et de l'Instrumentation Associ\'ee}\\
{\normalsize \bf CEA-DSM (Saclay)}\\
\vglue 1mm
{\normalsize \bf 
  R.~Azoulay, R.~Berthier, F.~Blondeau, N.~de~Botton, P.H.~Carton, M.~Cribier,
  F.~Desages, G.~Dispau, F.~Feinstein, P.~Galumian, Ph.~Goret, L.~Gosset, 
  J.F.~Gournay, D.~Lachartre,
  P.~Lamare, J.C.~Languillat, J.Ph.~Laugier, H.~Le~Provost, D.~Loiseau,
  S.~Loucatos, P. Magnier, J.~McNutt, P.~Mols, L.~Moscoso, P.~Perrin, 
  J.~Poinsignon,
  Y.~Sacquin, J.P.~Schuller, J.P.~Soirat, A.~Tabary, D.~Vignaud, D.~Vilanova.
}
\\[5mm]
{\large \bf Instituto de F\'{\i}sica Corpuscular}\\
{\normalsize \bf CSIC-Universitat de Val\`encia}\\
\vglue 1mm
{\normalsize \bf
  R.~Cases, J.J.~Hern\'andez, S.~Navas, J.~Velasco, J.~Z\'u\~niga.}         
\\[5mm]
{\large \bf Institut Fran\c{c}ais de Recherche pour\\
l'Exploitation de la MER}\\
\vglue 1mm
{\normalsize \bf J.F.~Drogou, D.~Festy, G.~Herrouin, L.~Lemoine, F.~Mazeas, P.~Valdy.}         
\\[5mm]
{\large \bf Institut Gassendi pour la Recherche\\Astronomique en Provence}\\
{\normalsize \bf INSU - CNRS}\\
\vglue 1mm
{\normalsize \bf Ph.~Amram, J.~Boulesteix, M.~Marcelin, A.~Mazure, R.~Triay.}
\\[5mm]
{\large \bf Oxford University}\\
{\normalsize \bf Physics Department}\\
\vglue 1mm
{\normalsize \bf
  D.~Bailey, S.~Biller, B.~Brooks, N.~Jelley, M.~Moorhead, D.~Wark.}         
\end{center}

\newpage
~
\setcounter{page}{1}
\pagenumbering{roman}
\tableofcontents

\newpage
\setcounter{page}{1}
\pagenumbering{arabic}
\addcontentsline{toc}{section}{\numberline{}Preamble - Introduction}
{\Large \bf Preamble}
\vglue 1cm

This document is mainly intended to describe the Astroparticle
aspect of ANTARES. The Sea Science part of the project is described
in \cite{in2p3}.
\vglue 1cm

{\Large \bf Introduction}
\vglue 1cm

We propose to observe High Energy Cosmic Neutrinos using a deep sea Cherenkov
detector. In what follows, we will elaborate on the potential
interest of such a study for Astrophysicists and Particle Physicists. For 
Oceanologists participating in the collaboration,
the main goal is a long term measurement of environmental parameters
in the deep sea. 

The physics detector principle is based on the detection of
Cherenkov light emission in either ice or water of upward going muons 
induced by
neutrino interactions in the medium surrounding the detector.

AMANDA \cite{prop-amanda} is a running experiment installed at the South Pole, 
for which it has been demonstrated that
deployment and logistics problems can be solved. It still remains to 
be demonstrated
that the quality of the ice allows an accurate measurement of the neutrino 
direction.

BAIKAL \cite{prop-baikal} is another running experiment installed in lake Baikal,
for which the feasibility as well as the capability to reconstruct
up going and down going muons have been shown. 
The relatively poor transparency of the
water and the shallow depth of the lake may be a limitation for 
further extensions.
Moreover, optical properties of deep sea water 
have been measured to be better than those of lake water.

DUMAND \cite{prop-dumand} was a pioneer for studies about deep ocean water detectors.
The funding of this project has been cancelled in 1996 by DOE \cite{sagenap}.

NESTOR \cite{prop-nestor} is a project planned to be installed in a deep sea 
site offshore from Pylos, Greece. 
Preliminary work started in 1989. The first step will be the
installation of
a few optical modules connected via an 
electro-optical cable to the shore and later of a complete tower.

In terms of depth, servicing and optical properties a deep-sea detector is
promising. We propose to explore the possibility
of a km-scale detector to be installed in a deep
site in the Mediterranean sea, for which a broad collaboration will be needed.
Furthermore, a variety of technical problems have to be solved.
Some of them are standard for particle physicists (choice of 
photo-multipliers, monitoring, trigger, electronics and data acquisition, 
analysis tools...), although the constraints coming from
the deep sea environment and the lack of accessibility
have to be fully taken into account. Others are more specific of sea science 
engineering, namely detector deployment
in deep water, data transmission through optical cables,
corrosion, bio-fouling of optical
modules, positioning. We have found technical support from collaborators
and partners
which have experience in this field (COM, CSTN, CTME, IFREMER, France 
T\'el\'ecom 
C\^ables, INSU-CNRS...).

We will test the sea engineering part of a detector including test deployments
close to the Toulon coast (France) where technical support is available and 
where several sites at depths down to 2500~m are easily accessible.
During the same time, issues connected to the accomplishment of a
large scale detector and the selection of an optimum site will be addressed.

We propose to build and install a demonstrator (a fully equipped 3-dimensional
test array) the design of which can be
extended to a km$^3$ scale detector. We plan to reach this goal within the next 
2 years.

\newpage
\section{Scientific motivation} 
The ANTARES project addresses several physics topics
which have in common the need for a long exposure of a large detector
shielded from charged cosmic rays.

In what follows, we will mainly describe the Astroparticle aspects of the
project. Only a few examples of possible Sea Science studies will be
mentioned.
\subsection{Scientific motivation in Particle Physics and Astrophysics}
\subsubsection{Introduction}
In Particle Physics, the unification of the four known forces has been and 
still is a major goal in the quest for our understanding of the Universe.
In the framework which is generally accepted, all the
forces are unified at very high energy. In order to have access to much higher 
energies than available at LHC in the near future, Big Bang 
relics or active cosmic objects can be used as providers of Ultra High Energy
particles (in excess of 10$^{20}$ eV, as already detected on Earth 
\cite{gaisser-book}). How Nature is able to accelerate
particles to energies far beyond human possibilities is still an unanswered
question. Taking advantage of
the available cosmic particles, the study of Ultra High Energies could help us,
on one hand, to test various models of acceleration mechanisms, and on
the other hand,
to constrain the different candidate theories which aim at extending the
Standard Model up to the Planck scale.

The supersymmetric extensions of the Standard Model \cite{jungman},
predict the existence of at least four neutralinos, linear combinations of
the two superpartners of the neutral SU(2)
gauge bosons (gauginos) and the two superpartners of the neutral Higgs particles
(higgsinos). In most of the models, the lightest neutralino is the Lightest
Supersymmetric Particle and thus is a candidate for dark matter. So, even if
superparticles were
discovered at accelerators, it would still be of major interest to constrain
both supersymmetric and cosmological models by observing
neutrinos produced in the annihilation of neutralinos remnants of the
Big Bang.

The atmospheric neutrino flux can also be studied for neutrino oscillations.

A High Energy Neutrino detector may
open a new observation window on the Universe, complementary to the photonic 
observations already in use, and help with the elucidation of the origin of 
the Ultra High Energy Cosmic Rays already observed.

\begin{figure}[p]
  \begin{center}
    \epsfig{file=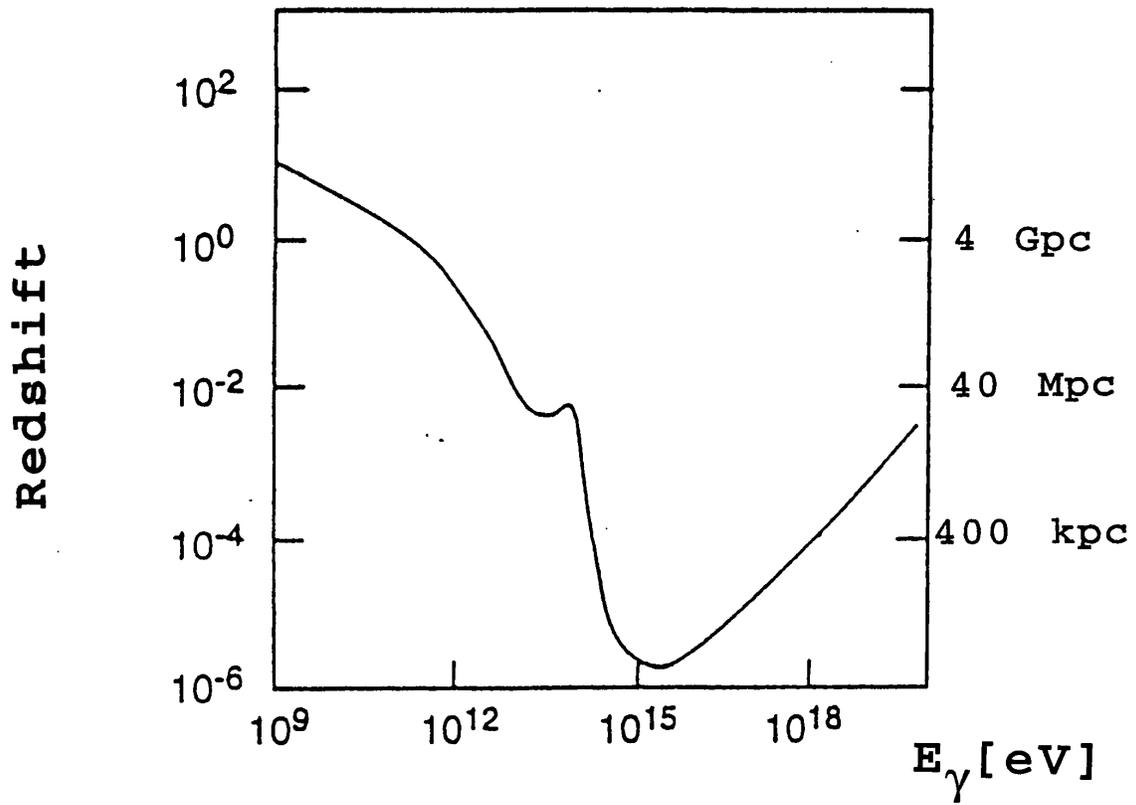,height=0.5\textheight}
    \caption{Attenuation length of cosmic photons as a function of their energy
    $E_\gamma$. The attenuation is due the interactions of these photons on
    the IR and microwave cosmic backgrounds.}
    \label{fig:CMB}
  \end{center}
\end{figure}

The use of neutrinos to observe the Universe has some intrinsic
advantages. Charged particles are sensitive to magnetic fields,
at the source, during their transport and in the Galaxy; so, except for those
with ultimate energies, they do not point to their emission source. In contrast,
neutrinos and photons are insensitive to those magnetic fields. However, high
energy photons are absorbed by a few hundred g.cm$^{-2}$,
when the interaction length of a 1 TeV neutrino is about 250~10$^9$ g.cm$^{-2}$.
Furthermore interactions of Very High Energy photons with the infrared
radiation and Cosmic
Microwave Background (2.7~K) limit their path length to distances smaller than
100 Mpc (see fig.~\ref{fig:CMB}).


Because they could originate from a common source, the combined study of both
the energy spectra of $\gamma$-rays and neutrinos emitted by cosmic sources is
necessary in order to tackle the question of the origin of the highest energy
cosmic rays and the nature of the mechanisms capable of producing them.

A variety of $\gamma$-ray detectors are already operating.
These detectors allow to explore the GeV region (satellites) and above
200\,GeV (large arrays and ground based Cherenkov imaging telescopes)
\cite{egret,esposito,whipple-1}.
These detectors have already shown evidence of point-like celestial
$\gamma$-ray sources.

As for cosmic neutrinos, a few examples of their detection exist at low
energies. Let us
mention the detection of the signals of solar neutrinos \cite{gallex},
below $\approx$ 10\,MeV, and of neutrinos from the supernova 
SN1987A \cite{sn1987},
at a few tens of MeV. It is, thus, of major interest to explore the
possibility to detect signals at higher energies. Several attempts have been
made with underground detectors, part of them devoted to proton decay
detection.
Due to the modest dimensions of such detectors ($\leq$ 1000 m$^2$),
 only upper
limits on the neutrino luminosities of several celestial bodies were obtained
\cite{underground,frejus}.
The expected fluxes of high energy neutrinos actually require a km-scale 
detector (see section~\ref{sec:rates}).
\subsubsection{High Energy Neutrino production}
\begin{figure}[p]
  \begin{center}
    \epsfig{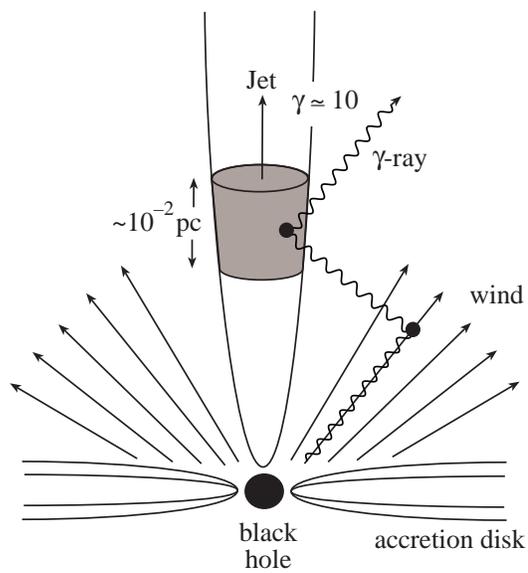}
    \caption{A possible model for the production of high energy photons and
    neutrinos in an AGN \protect\cite{halzen}.}
    \label{fig:agn-rep}
  \end{center}
\end{figure}

\begin{itemize}
\item{{\bf Neutrinos of cosmic acceleration origin:}\\
Point-like cosmic photon sources have been observed in the TeV range by the
ground based observatories \cite{whipple-1}.
Two mechanisms are possible for the production of these photons 
\cite{gaisser-book}.
The electromagnetic one is based on synchrotron radiation emitted by accelerated
plasma followed by inverse Compton scattering on electrons. 
The hadronic one is based on the decay of
$\pi^0$'s produced in hadronic interactions of accelerated nucleons
with a cosmic target (matter or photon field).

The first mechanism does not produce any neutrinos. The second one, 
in contrast,
gives rise to neutrino production coming from the decay of charged 
pions produced together with the neutral ones.

\[
  p/A + p/\gamma \longrightarrow
\begin{array}[t]{l}
 \pi^0 \\
 \downarrow \\
 \gamma + \gamma
\end{array} +
\begin{array}[t]{l}
 \pi^+ \\
 \downarrow \\
 \mu^+ + \nu_\mu \\
 \downarrow \\
 e^++\nu_e+\overline{\nu}_\mu
\end{array} +
\begin{array}[t]{l}
 \pi^- \\
 \downarrow \\
 \mu^- + \overline{\nu}_\mu \\
 \downarrow \\
 e^-+\overline{\nu}_e+\nu_\mu
\end{array} + ...
\]

Depending upon the precise characteristics of the cosmic beam dump and the
ratio of charged to neutral pions,
the $\nu / \gamma$ ratio, at a distance from the source, can take any value
between the $\nu / \gamma$ ratio at production (leaky source) and infinity
(shrouded source) \cite{gaisser-book}.

As the highest energy cosmic rays detected on Earth are very likely protons, 
it is unavoidable that cosmic neutrinos are created from their interactions
with matter and the Cosmic Radiation Background.

Possible {\bf galactic sources} are:
\begin{itemize}
  \item{{\bf X-ray binary systems}.
They are made of a compact star, such as a neutron star or a black hole, which
accretes the matter of its non compact companion. Strong
magnetic fields combined to plasma flows lead to a stochastic acceleration of
particles by resonant interaction with plasma waves in the magnetosphere of 
the compact star. The interaction of the
accelerated particles with the accreted matter or the companion itself
produces mesons which eventually decay into neutrinos.}
  \item{{\bf Young supernova remnants}. Protons inside supernova shells can
  be accelerated in the magnetosphere of the pulsar, by a first order Fermi
  mechanism in case of turbulence in the shell or at the front of the shock
  wave produced by the magneto-hydrodynamic wind in the shell \cite{berez}.
  The interaction of these protons with the matter of the shell gives rise
  to neutrino emission. The active neutrino phase lasts from 1 to 10 years 
  after the supernova explosion. Recent observations \cite{esposito} above
  100 MeV by the EGRET detector have found $\gamma-$ray signals associated with
  at least 2 supernova remnants (IC 443 and $\gamma$ Cygni).}
\end{itemize}

As for {\bf extra-galactic sources}:
\begin{itemize}
\item{{\bf Active Galactic Nuclei} (AGNs) are good candidates.
These galaxies are the brightest objects in the Universe. Their powerful and
compact central engine is thought to be made of a supermassive black hole
(10$^4$-10$^{10}$ solar masses). The energy powering the engine comes from the
accretion of the matter surrounding the black hole at a rate of a few solar
masses per year, leading to total luminosities in the range 10$^{42}$-10$^{48}$
erg/s~\cite{gaisser}. Different acceleration sites in the AGNs are envisioned
(see fig.~\ref{fig:agn-rep} and~\cite{halzen}), namely:
\begin{itemize}
  \item{close to the central engine,}
  \item{along  the radio jets,}
  \item{in hot spots terminating the jets in the radio lobes,}
\end{itemize}
leading to possible neutrino generation by interaction of accelerated protons
with surrounding matter in the central core (pp) or with dense photon fields 
in jets (p$\gamma$).
Emission of $\gamma$-rays up to $\approx$ 10\,GeV from Active Galactic Nuclei
have been well
established by EGRET \cite{egret}; two of them (Mrk 421 and Mrk 501)
have been also observed as 
emitters of high energy gamma rays (above 1\,TeV) by ground based observatories
\cite{whipple-1}.}
\item{We may include in the potential extra-galactic
sources {\bf Gamma Ray Bursts} although it cannot
yet be excluded they might be located in the
extended Galactic halo.
GRB's have been observed for many years and
recent detectors observed
bursts at a rate of one per day. Their location was first thought
to be galactic, now extra-galactic (cosmological)
models are strongly favored \cite{Cosmol}.
GRB's are expected to emit a big part of their
energy in neutrinos \cite{Pacz,Plaga}.
Recently, emission of neutrinos of very high energies 
from GRB fireballs
was predicted \cite{Waxman97}. The observation of 
gamma, X, optical and radio signals from GRB 970228
and GRB 970508 \cite{SAX}
(estimated to be at z $\geq 0.85$ ) is in agreement
with this model.}
\end{itemize}

By their very existence, {\bf galactic and extra-galactic cosmic rays}
guarantee the production of high energy cosmic neutrinos. Indeed, the primary
cosmic rays can interact with:
\begin{itemize}
  \item{the production medium,}
  \item{the Cosmic Microwave Background (2.7~K) \cite{yoshida},}
  \item{interstellar gas in our galaxy \cite{gaisser},}
  \item{the Earth atmosphere \cite{volkova},}
\end{itemize}
to produce mesons which eventually decay into neutrinos.}
\item{{\bf Neutrinos of non-acceleration origin:}\\
Topological defects (cosmic strings, monopoles) \cite{berez-primord} and non
baryonic dark matter \cite{gaisser} bring into play very heavy entities which
by quantum evaporation, collapse or annihilation eventually give rise to an
emission of high energy neutrinos.
\begin{itemize}
\item{Monopoles and cosmic strings are topological defects 
likely to have been formed in the symmetry breaking phase
transitions that occured in the early Universe. Inside these defects,
the vacuum expectation value is 0 while everywhere else 
it is of the order of the symmetry breaking scale 
($\sim 10^{16}$ GeV). These defects are stable but can be destroyed by
collapse or annihilation \cite{bhatta}, releasing the energy trapped in them
in the form of massive quanta $X$ which eventually decay into hadrons,
leptons, photons and neutrinos with energies up to $m_X \sim 10^{16}$ GeV.
The rate of release of $X$ particles is given by:
\[
    \frac{dN_X}{dt} = \kappa m^p_X t^{-4+p}
\]
where $t$ is the Hubble time. $\kappa$ and $p$ are dimensionless constants
the value of which depends on the characteristics of the topological defects:
\begin{itemize}
  \item $p = 0$ for saturated superconducting cosmic string loops,
  \item $p = 1$ for collapsing cosmic string loops and annihilating 
  monopole-antimonopole bound states.
\end{itemize}}
\item{Neutralinos, remnants from the Big Bang, 
move in the halo of the Galaxy with velocity of a few hundred km/s.
They can encounter celestial bodies, lose energy by elastic scattering
on the nuclei these bodies are made of, and stay trapped in them. This results
in a high concentration of neutralinos in the Sun and the Earth which 
enhances their annihilation rate per unit volume giving rise, amongst
others, to a neutrino emission~\cite{jungman}.}
\end{itemize}
}
\end{itemize}
\subsubsection{Detection of High Energy Cosmic Neutrinos}
\begin{figure}[p]
  \begin{center}
    \mbox{\epsfig{file=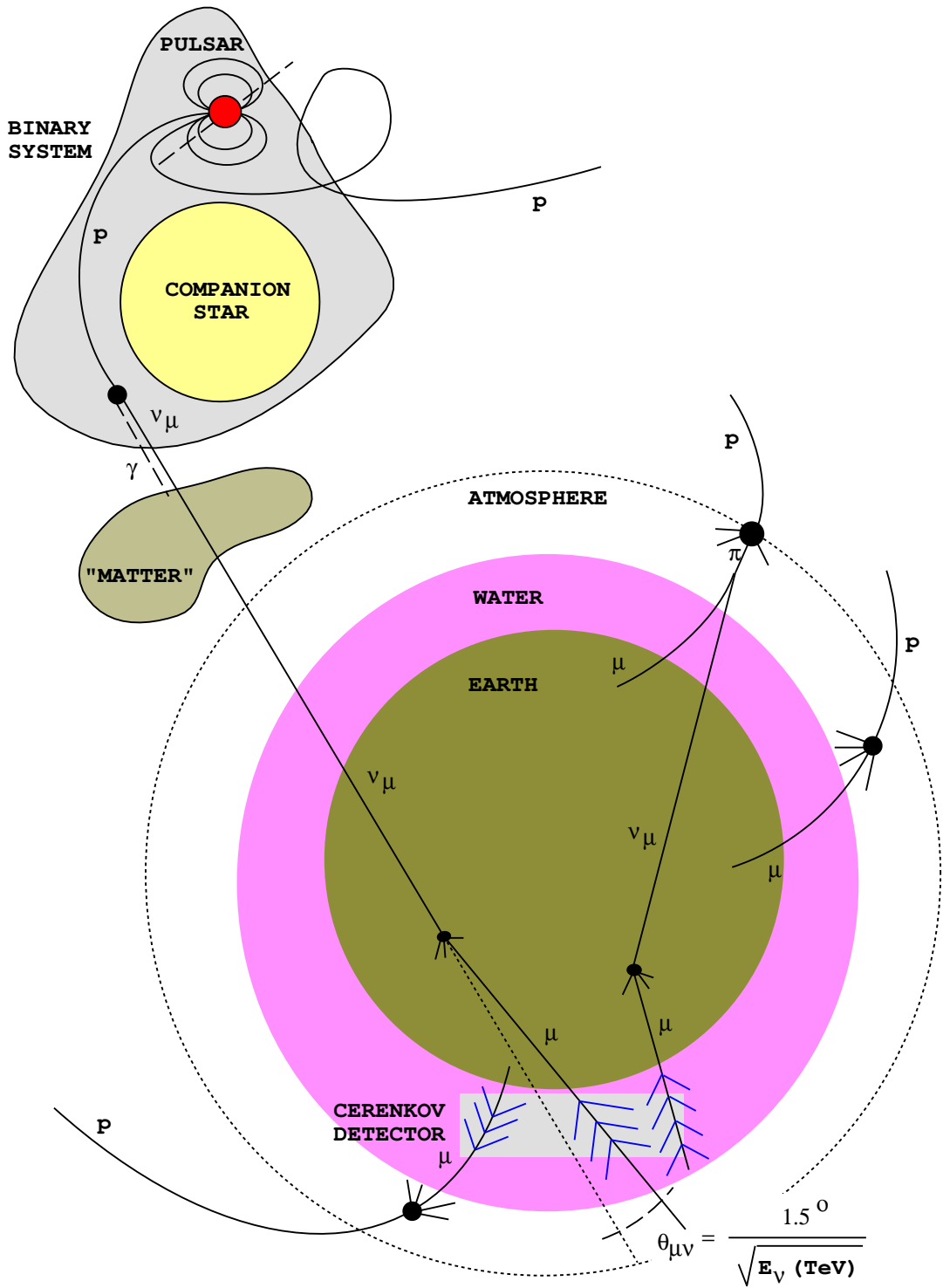,height=0.9\textheight}}
    \caption{Neutrino detection principle}
    \label{pulsar}
  \end{center}
\end{figure}

\begin{figure}[p]
  \begin{center}
    \mbox{\epsfig{file=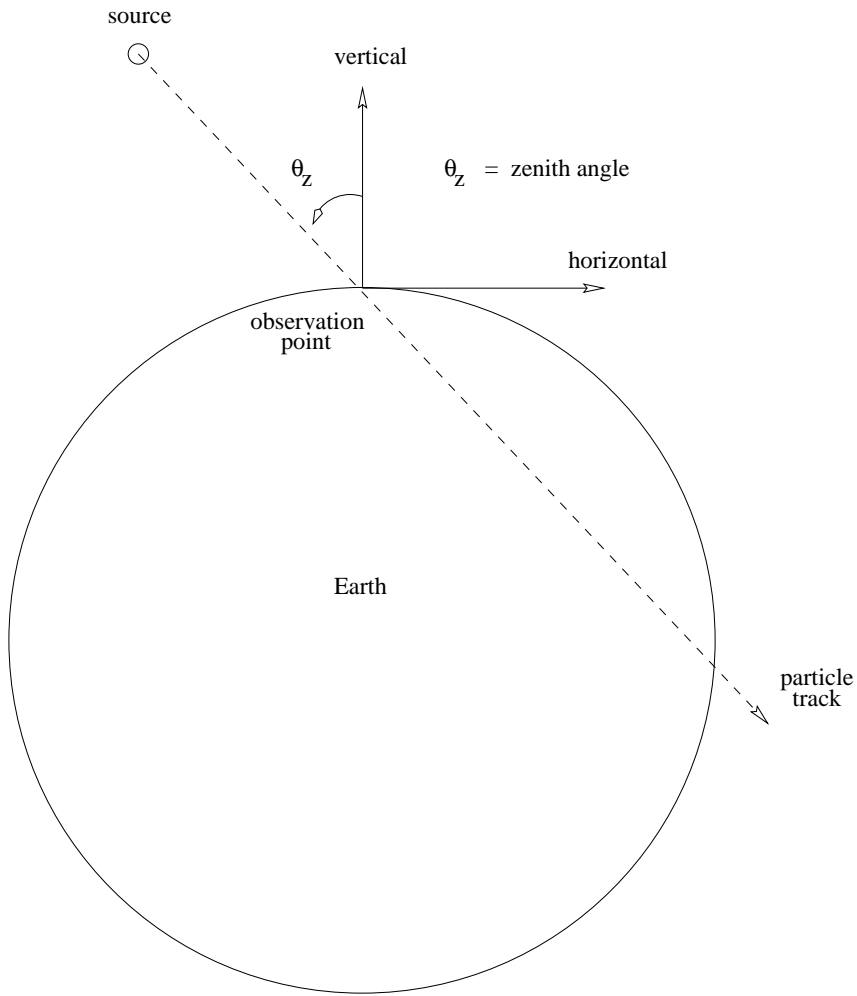,height=0.6\textheight}}
    \caption{Definition of the zenith angle. The zenith angle of an incoming
    particle is the zenith angle of the source it originated from.}
    \label{fig:nadir}
  \end{center}
\end{figure}

\begin{figure}[p]
  \begin{center}
  \mbox{\epsfig{file=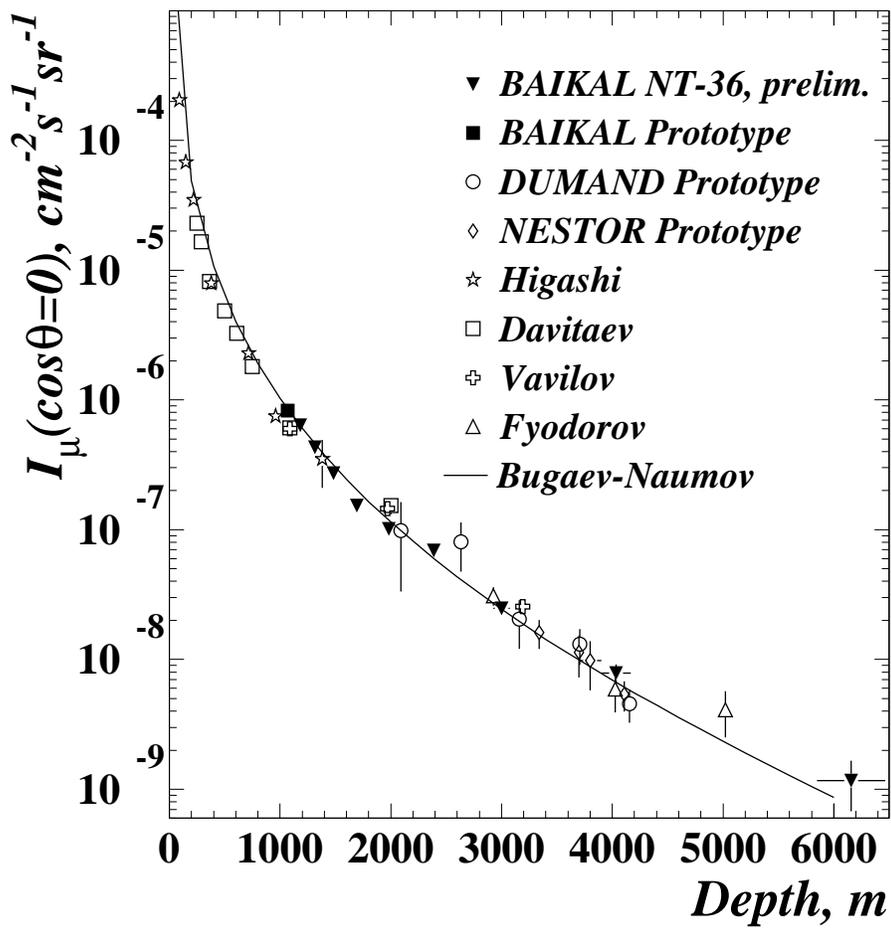,width=0.9\textwidth}}
  \caption{Vertical atmospheric muon flux as a function of depth in meters
  water equivalent \protect\cite{muflux}.}
  \label{fig:flux-mu}
  \end{center}
\end{figure}

\begin{figure}[p]
  \begin{center}
  \mbox{\epsfig{file=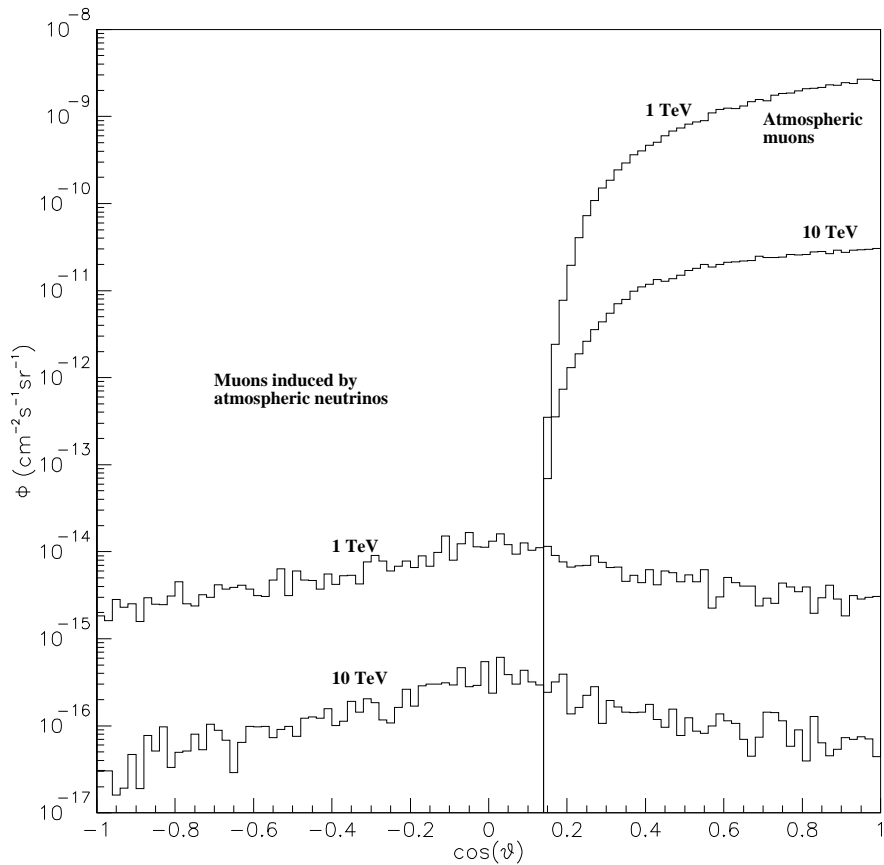,width=0.9\textwidth}}
  \caption{Atmospheric muon flux (under 2\,300\,m of water)
   and atmospheric neutrino induced muon flux as a function of the zenith
  angle for two muon energy thresholds (1 and 10 TeV).}
  \label{fig:zenith}
  \end{center}
\end{figure}

\begin{itemize}
\item{{\bf Detection principle}:\\
All potential sources produce $\nu_e$'s as well as $\nu_\mu$'s.
One could detect $\nu_e$ interactions by observing electromagnetic and hadronic
showers of contained events. However, at this stage, we will only concentrate
on $\nu_\mu$ detection.

High energy muon neutrinos can be detected by observing
long-range muons produced in
charged current neutrino-nucleon interactions with matter surrounding the
detector (see figure~\ref{pulsar}). To reduce the background from direct
muons produced in the atmosphere, the neutrino telescope should be located at a
depth of several kilometers water equivalent and one should only consider
muons with a
zenith angle greater than $\sim$ 80$^\circ$ ({\it i.e.} $\sim$ 10$^\circ$ above 
the horizon (see fig.~\ref{fig:nadir}), the value for which the flux of
direct atmospheric muons becomes smaller than the flux of atmospheric neutrino
induced muons under 3000 meters of water.
At high energy, the outgoing muon is emitted
in the same direction as the incident neutrino 
($\theta_{\mu\nu}\approx 1.5^{\circ}/\sqrt{E[TeV]}$) allowing to point back
to the source of the neutrino emission.


When passing through sea water, the muon emits Cherenkov light which is 
detected by a three-dimensional matrix of photo-multiplier tubes. The
measurement of the arrival time of the Cherenkov light on the 
photo-multiplier tubes allows the reconstruction of the muon direction. 
The amount of light allows to estimate the muon energy giving a lower limit 
on the energy of the parent neutrino.
}

\item{{\bf Neutrino-matter interaction}:\\
  In order to calculate the flux of muons going through such a detection
  set-up, we need to know, besides the $\nu_\mu$ and $\bar\nu_\mu$ fluxes, the
  neutrino-nucleus interaction cross-section, the attenuation of the neutrino
  flux in the Earth and the range of the induced muon. Details of the
  calculation can be found in the appendix.
}
\item{{\bf Physical backgrounds}:\\
  Physical backgrounds to cosmic neutrinos essentially come from neutrinos and
  muons produced in atmospheric showers resulting from the interaction of primary
  cosmic rays with the Earth atmosphere.
  \begin{itemize}
    \item Cosmic neutrinos cannot be distinguished from atmospheric neutrinos
    which originate from the decay of charged pions and kaons produced by 
    primary cosmic rays
    interacting with the Earth atmosphere. An additional ``prompt'' 
    contribution arises from heavy flavor
    production and decay; this contribution is small compared to the
    $\pi$-$K$ one for $E_\nu < 10^5$ GeV.\\
    Atmospheric and cosmic neutrino fluxes $\Phi_\nu$ can be approximated by simple
    power laws:
\[
   \frac{d\Phi_\nu}{dE_\nu} \propto E^{-\gamma}_\nu
\]
    The flux of atmospheric neutrinos has a $\gamma \simeq$ 3.7 \cite{gaisser}.
    However for
    cosmic sources we expect $\gamma$ to be around 2. The greater
    slope of the atmospheric neutrino spectrum makes the signal/background
    ratio improve with energy.
    \item Direct isolated muons and muon bundles are also produced copiously by
    interaction of primary cosmic rays with the Earth atmosphere. For small
    zenith angles, this background is several orders of magnitude
    higher than the atmospheric neutrino induced muon background. So, it is
    necessary to shield the detector from these direct muons in order to lower
    their detection rate and thus the probability of mis-reconstructing
    an atmospheric muon or a muon bundle as an upward going neutrino-induced
    muon. The background due to downward going muons backscattering upwards is
    negligible compared to the atmospheric neutrino induced muon background.
  \end{itemize}
  
  Fig. \ref{fig:flux-mu} shows the atmospheric muon vertical flux as a
  function of depth in meters of water~\cite{muflux}. Fig. \ref{fig:zenith} 
  shows the
  atmospheric muon flux 
  (under 2\,300\,m of water) compared to the atmospheric
  neutrino induced muon flux as a function of the zenith angle $\theta_z$ for
  two muon energy thresholds (1 and 10 TeV).
%
%
%
}
\end{itemize}  
\subsubsection{Fluxes and Rates}
\label{sec:rates}
\begin{figure}[hhh]
  \begin{center}
    \mbox{\epsfig{file=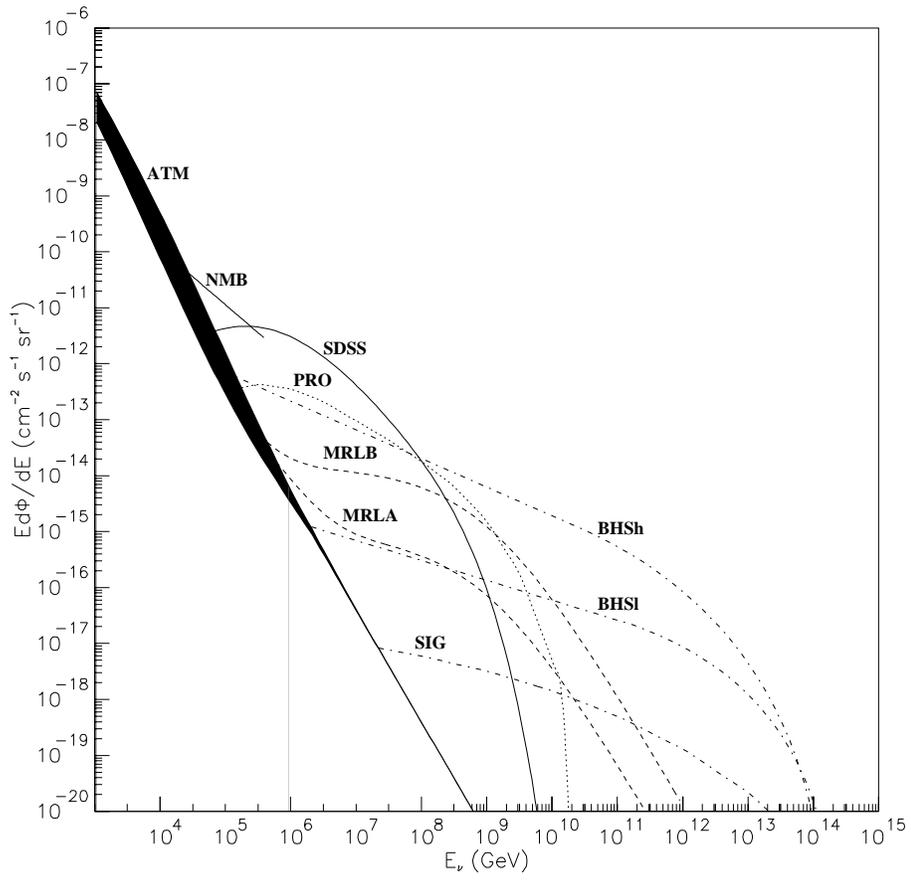,width=0.9\textwidth}}
    \caption{Neutrino fluxes at Earth from different sources (see text)}
    \label{fig:agn-flux}
  \end{center}
\end{figure}

\begin{itemize}                       
\item{{\bf Galactic sources}:\\

It is generally assumed that the neutrino flux emitted by a galactic
source is proportional to 
$E_{\nu}^{-\gamma}$
($\gamma \approx$ 2 from $\gamma-$ray observations above 10~GeV.)

In order to calculate the sensitivity of a detector to a given proton
luminosity ${\cal L}_p$, the number of expected
muons with energies above a given energy threshold during the exposure time has
to be compared with the background due to atmospheric neutrinos.

This calculation
has been performed with a Monte~Carlo program
taking into account the $\nu-N$ interaction (see~\cite{quigg,botts} and the 
appendix)
and the muon
propagation in the medium surrounding the detector. The expected number of
muons is given by:
\[ N_S =\Phi_0 \left( \frac{1\,{\rm kpc}}{D} \right)^2
         \left( \frac{\epsilon}
         {0.1} \right) \left( \frac{{\cal L}_p}{10^{38}{\rm erg\,s}^{-1}}
         \right) ST \]
where $\epsilon$ is the ratio of the total neutrino luminosity to the
total proton luminosity. The value of the neutrino induced muon 
flux $\Phi_0$ is
given in table~\ref{lmtb:tab1} for a differential
spectral index $\gamma = 2.2$ (to be conservative) and for different
muon energy thresholds. The detector exposure $ST$ is a convolution of the
effective area $S$, which is supposed to be independent of the muon direction,
with the running time $T$, calculated taking into account the on source
duty factor (1~km$^2$yr $\simeq$ 3~10$^{17}$ cm$^2$s).


The fluxes of muons induced by atmospheric neutrinos, averaged over the
detectable hemis\-phere, have been calculated using the neutrino flux
given in ref.~\cite{volkova} and
are also given in the last column  of
table~\ref{lmtb:tab1}. This flux allows us to calculate
the background which contaminates the signal.

The proton luminosity ${\cal L_p}$ that can be detected as a signal,
exceeding by
5 standard 
deviations after a year
the atmospheric neutrinos background, is given in table~\ref{lmtb:tab2} 
(the minimum signal considered is always larger than 10 events a year).

%

This table clearly shows that
better sensitivities can be obtained for the highest values of the muon energy
threshold. So, a cubic kilometer detector is particularly well aimed at the
detection of sources with total proton luminosities of
10$^{34}$-10$^{35}$\,erg\,s$^{-1}$ at distances of $\sim$\,1\,kpc and
between
10$^{36}$-10$^{37}$\,erg\,s$^{-1}$ for the whole Galaxy
($\simeq$\,10\,kpc).

For individual known sources, the calculation of the detectable
luminosity can be
performed by taking into account the distance of each individual source, the
fraction of time ${\epsilon}_t$ during which the source is below the horizon,
the latitude of the detector
and the flux of the background atmospheric neutrinos averaged along the apparent
path of the source. As an example,
these values are given in table~\ref{lmtb:tab3} for a
detector of 1\,km$^2$
located at a latitude of $45^{\circ}$ running for one year
and for a threshold in energy of
1\,TeV.

%
}
\item{{\bf Extra-galactic sources}:\\

The models of AGN generated neutrino fluxes differ in their production sites and
production mechanisms. We consider three types of models which do not
contradict existing measurements ({\it e.g.} the Frejus upper limit at 2.6
TeV \cite{frejus}):
\begin{itemize}
  \item{SDSS \cite{stecker} which has significant contributions from pp and
  p$\gamma$ interactions in the accretion disk,}
  \item{NMB \cite{nellen} for which pp interactions are the dominant
  source of neutrinos,}
  \item{PRO \cite{prothe-blazar}, MRLA and MRLB \cite{mannheim-blazar}
  for which are taken into account the contribution of neutrinos generated 
  in blazar jets
  (radio-quasars for which the jet axis is more or less aligned with the line of
  sight).}
\end{itemize}

The corresponding diffuse fluxes of muon neutrinos 
for these models is shown in fig.~\ref{fig:agn-flux} together with the
angle-averaged atmospheric neutrino flux (ATM)~\cite{volkova}.

The rates of neutrino induced muons are calculated from these fluxes using the CTEQ3-DIS parton
distribution functions and a Monte Carlo propagation of the muons.
The results for a 1\,km$^2$ effective area 
detector are summarized in
table~\ref{lmtb:tab4}, for muon energy thresholds E$^{min}_\mu$ ranging from
1 TeV to 100 PeV. Above 10\,TeV the event rates
predicted for neutrinos coming from AGNs become larger
than for atmospheric neutrinos.\\~~~\\
If some of the AGNs are powerful enough sources, they could be detected
individually.
A possible method to estimate the fluxes from individual AGNs is to select all
AGN sources of the Second EGRET Catalog and to assume that the neutrino flux is
equal to the gamma-ray flux. 
We have extracted from this
catalog all AGNs for which the gamma-ray flux between 100\,MeV and 10\,GeV
and the spectral
index $\gamma$ were measured, and
we have
extrapolated the flux above 1\,TeV. These values were assumed as neutrino
flux above this threshold.
The basic assumption for this is that all gamma-rays
are of hadronic origin. In the case where an important fraction of gamma-rays
are of electromagnetic origin the values that we quote for neutrinos must be
scaled down. On the other hand, for a given ratio neutrino/gamma-ray flux at
the production point, one could expect an increase of this ratio at the
detection point because the absorption effect is more important for gamma-rays
than for neutrinos. 

The most critical parameter from the point of view of the uncertainty of the
extrapolation is the spectral index. The measured values of this parameter are
fairly dispersed. So we have performed the calculations by using three 
different methods. The first method consists in using the values of $\gamma$ as
they are reported in the EGRET Catalog. For the second method, we have assumed
that the dispersion on the different values of $\gamma$ is due to uncertainties
and we use
$\gamma = 2$ which is the value generally used. 
The third method uses an average value of $\gamma$
obtained by using weighting factors equal to the inverse squared of the
experimental error on $\gamma$. For this latter method, we used the 24
measurements reported on the Second EGRET Catalog and the measurement
reported in
ref.~\cite{hartman} for 3C\,279. We obtain in this way 
${\gamma}_{\mathrm av} = 2.12 \pm 0.03$  
(with a fairly poor chi-square value of 72.7 for 24 degrees of freedom).
 
There were 4 sources with a declination bigger than 45$^{\circ}$ which have been
discarded because they cannot be detected by a detector which is located at
45$^{\circ}$N. For the remaining 21 sources, the neutrino fluxes have been
corrected to take into account of the fraction of time (${\epsilon}_t$) during
which
the source is below the horizon.
 
Table~\ref{kn:agn_discr} gives for each one of the remaining 21
sources, its name in the Second EGRET Catalog, its right ascension (RA), its
declination ($\delta$), the
spectral index ($\gamma$), the ${\epsilon}_t$ value
 and the three estimates of the numbers of muons
induced by neutrinos with muon energy above 1\,TeV expected per km$^2$ and per
year. 
Table~\ref{kn:agn_discr} shows that 
due to the low background level ($\approx3 \times {\epsilon}_t$ per year and per
km$^2$ for an angular cut of 1$^{\circ}$),
for each one of the three different methods of calculation
there are several sources for which a statistically significant signal could be
detected in one year with a km$^2$ detector.

From Gamma Ray Burst sources,
in \cite{Waxman97} the rate of upcoming muons from neutrinos 
above $\simeq$ 100 TeV
in an underwater detector is expected to be between 10 and 100 /km$^2$/year.
These neutrinos would be distinguished from the atmospheric ones
due to their correlation to GRB's, which can be precise thanks
to the brief emission interval of the gammas.
}
\item{{\bf Neutrinos from topological defects}:\\

We considered three models: BHSl and BHSh  
(with $m_X = 10^{15}$ GeV, $p = 1.5$ and $0.5$ respectively) \cite{bhatta}
and SIG (with $m_X = 2~10^{16}$ GeV, $p = 1$, constrained by the
1-10 GeV $\gamma$-ray observational data assuming an extra-galactic magnetic
field of $10^{-12}$ G) \cite{sigl}. The corresponding expected diffuse
neutrino fluxes are displayed in fig. \ref{fig:agn-flux} and the corresponding
induced muon rates for a km$^2$ effective area detector are shown in table 
\ref{lmtb:tab4}. Only the most optimistic amongst these three models (BHSh)
gives rise to a
detectable signal in one year with a km$^2$ detector.
}
\item{{\bf Neutrinos from dark matter}:\\

%
%

The signal will consist of an excess of neutrino flux coming from the Sun
or from the center of the Earth.

The calculation of the sensitivity of a detector depends on
the parameters of the theoretical model and the atmospheric neutrinos
background.

This calculation has been performed in~\cite{jungman,gaisser,bottino} in terms 
of the sensitive
area required to detect a 4 standard deviations signal as a function of the
neutralino mass. The results of this calculation shows (see  
fig.~\ref{fig:wimps}) that a detector with an area of 1\,km$^2$
running for one year would
be sensitive to a range of neutralino masses extending up to a few TeV.
}                  
\item{{\bf Atmospheric neutrinos}:\\

For all subjects described above, the atmospheric neutrinos constitute a 
source of
background which is suppressed by using an energy threshold of $\approx$\,10\,
TeV. Nevertheless, atmospheric neutrinos may be an interesting subject
of study by lowering the energy threshold to $\approx$\,10\,GeV.

In this case, the analysis of the angular distribution of muon neutrinos 
will allow
the study of the flux as a function of the thickness of matter crossed
through the Earth
which varies from a few tens of km to 13\,000\,km depending on the zenith
angle. This should allow to look for neutrino oscillations in a domain of the
oscillation parameter space where Kamiokande has shown some evidence of such
an effect \cite{Kamioscil}. However, even if
atmospheric 
neutrino oscillations were confirmed by SuperKamiokande \cite{superK},
it would be useful to
cross-check such a result using a different technique.
}
\end{itemize}
\subsection{Scientific motivation in Sea Science and Geology}

Several fields of science are interested in long term measurement in
deep sea environment, some need real time information (e.g. seismology), some do
not and there are many efforts going on in this direction.

At the time being, the foreseen programme deals mainly with the carbon cycle,
with current measurements and with geological measurements.

It is certainly not a complete list of what can be done and future
developments may enlarge the scope of these studies.
\subsection{The km-scale detector}
In order to observe with relevant statistics the diffuse flux of neutrinos from
AGNs and from point sources, and to be sensitive to
neutralinos with masses above 1 TeV, a
km-scale detector is necessary. We believe that the
technology needed to build such a detector is available. The first phase of
the ANTARES project, described below, aims at demonstrating the feasibility of
such a detector.

\newpage 
\section{Neutrino telescope concept} 
\subsection{Basic principle}

The detector consists of a 3-dimensional array of optical modules (OMs)
immersed in the deep sea, 100~m above the sea bed.
Figure~\ref{trames} shows two possible arrays, the first one with a
limited number of OMs will have a higher muon energy threshold
than the second one which has more OMs. An OM is made of a
photo-multiplier tube (PMT), its electronics and power supply housed in a
pressure resistant glass sphere. We have considered 8 and 15 inch PMTs
with hemispherical
photo-cathode. Each PMT is shielded against the Earth
magnetic field by a high permittivity metallic cage.
                                                               


The OMs are hooked to a mechanical structure which ensures the
stability of their relative positions in water.

A muon neutrino converts to a muon via a charged current interaction with a
nucleon of the rock or the 
surrounding sea water. {\vC} light is
emitted in the sea water by the muon and then is detected by the OMs. The
measurement of the arrival time of the light over at least five OMs allows the
reconstruction of the muon direction. The amount of collected light
allows to estimate
the muon energy which is a lower limit of the neutrino energy.

Optical beacons consisting of glass spheres housing blue GaN LEDs allow a local
time and amplitude calibration of the OMs. However, to be able to perform a
global time calibration of a large scale detector, a light source such as 
a YAG laser is required, because of its greater light yield.

There are different ways of arranging physically the photo-multipliers
(orientation, pairing\ldots). A final choice will result from optimization of
efficiency, resolution and cost.

Different schemes for the data transmission can be
considered \cite{LBL-38321,capone-nestor}. For the readout of the
optical modules one can
have a cluster of about 10 OMs
grouped around a local controller. Most of the signals that an OM will detect
will consist of single photoelectrons (SPEs). These SPEs are mostly caused by
light emitting background processes such as beta decays of $^{40}$K or 
bioluminescence. For $^{40}$K the typical
counting rates have been measured 
(see optical background measurements in section \ref{sec:optmes})
to be about 60 kHz (20 kHz) at the 0.5 (0.3) PE level with a 
15$^{\prime\prime}$ (8$^{\prime\prime}$) 
photo-cathode Hamamatsu PMT.
In order
to reduce this rate to about 1 kHz, PMT signals can be sent to the controller
which elaborates a local coincidence between neighboring PMTs and triggers 
the digitization of the
signals. The digital information is transmitted through several network
nodes in the array to a terminal node. With such a scheme using front-end
trigger and digitization, the data flow of a km-scale detector can be
handled with conventional techniques such as coaxial cables and Ethernet
protocol. The terminal node is connected to an electro-optical cable bringing
the power and slow control commands from the shore. In the terminal node the
digital information modulates the light output of laser diodes. This light is
then transmitted to the
shore via optical fibres.

\subsection{Status of other projects}

\begin{itemize}
\item {{\bf AMANDA}:\\

The AMANDA collaboration intends to build a neutrino telescope in the
Antarctic ice cap \cite{prop-amanda}. Strings of OMs are buried in
holes drilled in the ice.

The AMANDA collaboration has already deployed four detector strings
at a depth of 1 km at the South Pole; during the 1995-1996 and
1996-1997 campaigns, they
extended the depth to 2~km with an overall number of optical modules greater
than three hundred. The analog signals
are sent to the surface via coaxial cables (or twisted pair cables), which
also supply the high voltage.

Compared with a deep-ocean site, the ice provides a stable platform for safe
deployment as well as a natural mechanical support for the detectors. The
counting room can be installed at the surface above the detector thus 
simplifying the power and data transmission. The ice is a very pure and
transparent medium. It is also free of radioactive elements such as
$^{40}$K as well as bioluminescence.

However the ice medium has some drawbacks. Impurities such as air bubbles
are trapped in the ice crystal lattice, causing the light to scatter. After
several scatterings, the direction of the photons is randomized isotropically.
The measured scattering length $\Lambda = \lambda/(1-<\cos\theta>)$ ($\lambda$
is the scattering length between two diffusions, $\theta$ is the scattered
angle) ranges from 20~cm at 1000~m depth to 25~m (see for example 
\cite{price})
at 2000~m depth. The
depth in the ice is limited to about 2500~m while one may go deeper in the
ocean. Due to its location at the South Pole, only the northern sky can be
observed.}

\item {{\bf BAIKAL}:\\

The BAIKAL collaboration has deployed an array of 96~OMs at 1~km depth in
Lake Baikal \cite{prop-baikal,ISVHECRI}.  The deployment is performed
in winter from the frozen
surface of the lake. In order to suppress a background rate of tens of kHz in
each OM from $^{40}$K and bioluminescence in the lake, they pair OMs and
look only at coincidences. The counting rate is reduced to a few hundred
Hz. Half the OMs point upwards to achieve the same acceptance over the
upper and lower hemispheres.

The array of 36~OMs has been operating
since April 1993, recording more than 6.5\,10$^7$ muons in one year.
They have reached 
a record up/down rejection
ratio of 10$^{-4}$ and, according to their Monte Carlo, will reach
10$^{-6}$ (atmospheric neutrino level) when they deploy their full complement
of 200~OMs. With their 96 OMs, they should identify 1 to 2 upward
neutrinos events per week.}

\item {{\bf DUMAND}:\\

DUMAND has been a precursor in the field and they have studied many aspects
of the problems \cite{prop-dumand}. A lot of their experience has
to be taken into account.

A recent SAGENAP report \cite{sagenap} has recommended DUMAND to be terminated.}

\item {{\bf NESTOR}:\\

NESTOR is planned to be installed at a depth of about 3.8~km depth in the 
Mediterranean sea \cite{prop-nestor}.
At this
depth, the down-going muon flux is about 10$^4$ times the up-going muon flux
from atmospheric neutrinos, while it is 10$^5$~-~10$^6$ for AMANDA and BAIKAL. 
Half of its optical modules will point upwards. NESTOR is
designed to have a higher number density of OMs in the central volume, to
enable local coincidences on lower energy events.

The mechanical structure of NESTOR is a 12~floor tower, each floor composed
of a titanium star supporting a hexagonal array of OMs at the ends. Electronics
is housed in a central titanium sphere on each floor. FADCs and
memory operating at 300~MHz digitize the OMs signals and digital data are
transmitted to the shore via a 12~fibre electro-optical cable. Each fibre is
connected to a floor.

Since 1989, many tests have been performed. The first tower 
is expected in the near future. A further phase includes the deployment
of seven towers.}

\end{itemize}

Presently, both the AMANDA and BAIKAL experiments have demonstrated their 
ability to deploy optical modules using ice as a support. However, better 
optical properties of the transparent medium and a deeper
installation justify the need for a deep sea detector.

\newpage
\section{R \& D Programme of ANTARES} 
       
The feasibility of the deployment and operation of a large
detector in the deep sea needs to be demonstrated. Scientists of other fields
have deployed acoustic
detectors connected to the shore and used them reliably for many years. We
intend to built a demons\-trator which will prove the feasibility and 
will allow us to estimate the cost of a large scale detector.
In order to ensure that the exercise is meaningful, we must only use
techniques that can be extrapolated to a km-scale detector.\\
\vglue 0.5mm
The ANTARES programme can be divided into 3 stages, which will be discussed
in detail below:

\begin{enumerate}
\item{deployment of simple strings (these operations have already started);}
\item{deployment of a string connected to the shore via an
      electro-optical cable;}
\item{deployment of several strings and their submarine electric
interconnection (demonstrator).}
\end{enumerate}

For the completion of these stages, we have asked for the support of experts. 
We have started a 
collaboration with IFREMER and Centre d'Oc\'eanologie de Marseille
(INSU) as well as CSTN, France T\'el\'ecom C\^ables, CTME\ldots

More specifically, not only the know-how is requested, but also the
availability of ships and
submarines (IFREMER owned submarines: Cyana, Nautile and 
ROV~6000;
dynamical positioning ships which are needed should be available).

The optical module, the data transmission, the trigger and
the slow control of the expe\-riment are under study. 
In general, quality assurance 
will be pursued.

The geometrical arrangement of the array of optical modules needs to be
studied in order to optimize the effective area of the detector, the
angular resolution and the rejection against down-going muons.
                                                               
Site studies including water transparency, optical background and bio-fouling
measurements have already been started and need to be continued.
All the tools for site tests are either completed or in construction
and in this first phase of operation we have started to learn:
\begin{itemize}
\item how to build strings of detectors,
\item how to deploy and recover them,
\item how to handle difficulties associated with deep sea operations.
\end{itemize}

\subsection{Site studies}

\subsubsection{Quality of a site and measurements}
\label{sec:optmes}

Many different criteria have to be looked at before choosing a site for the
deployment of a large-scale detector:

\begin{enumerate}
\item{Water quality:\\
      The sensitivity of the experiment will depend on the water
      transparency and the light scattering at large angles. The
      knowledge of the dependence of these two parameters as a function of the
      wavelength is required. These measurements are delicate as they need
      to be made in-situ
      with a 30m long measurement system with well characterized
      optical sources at different wavelengths.
       Some measurements by DUMAND and NESTOR exist 
      (figure~\ref{attenuation}). A sketch of a design of our water transparency
      measuring device is shown in figure~\ref{fig:tests}a.
%
%
      }
\item{Water depth:\\
      The water above the detector is a natural shield against atmospheric
      muons. The rate of down-going muons drops by a factor 10 going from 
      2500 to 4400~m.
      Figure~\ref{fig:zenith} shows the rate of muons induced by atmospheric
      neutrinos and of direct atmospheric muons for an undersea depth of
      3000~m.}
\item{Optical background:\\
      $^{40}$K dissolved in salt water decays emitting electrons with an
      energy spectrum up to 1.3 MeV. Each electron produces 5 Cherenkov photons
      on average in the wavelength sensitivity window of the PMT. This gives
      rise to a photon flux of about 100\,cm$^{-2}$s$^{-1}$ for a 50~m water
      attenuation length. This light emission is very likely to be site
      independent in the Mediterranean. On the other hand, bioluminescence 
      (light emitted by a wide range of sea animal species) is time dependent
      and also site dependent.      
      Bioluminescence in the  deep sea is not well 
      known and one should measure
      the time structure of the emitted light as well as spatial correlations.
      Those measurements are currently under way, see 
      figure~\ref{fig:tests}b. 
      The first
      measurements have been done in October 1996 and in January 1997
      at a depth of 2400 m
      about 25 km off-shore from Toulon (see 42$^\circ$50'N-6$^\circ$10'E in
      fig.~\ref{carte-toulon}).}
\item{Bio-fouling:\\
      Deep sea bacteria have the tendency to colonize the surface of 
      immersed objects where they gather to form
      a sticky bio-film and make sediments hold on to it. This process is
      called bio-fouling. 
      The speed of formation of the bio-fouling is site
      dependent. It may affect the 
      transparency of the
      optical module in the long term. The rate of accumulation has to be 
      measured at the
      site and anti-fouling solutions have to be investigated. Sea science
      physicists are proposing different practical ways. Collaborations with
      EEC partners of the Bio-film Reduction on Optical Surfaces
      programme has started and measurements using the line described in  
      figure~\ref{fig:tests}c are under way.}
\item{Deep sea currents:\\
      Currents have to be taken into account in the
      mechanical design of the detector. Currents are changing during 
      the year, so they have to be measured on a year scale over the 
      vertical range
      covered by a future detector. This will be done in collaboration with
      sea science specialists.\\
      Benthic storms can generate very high currents. The rate at which 
      they have occured in the last 500 years can be deduced
      by geologists from a campaign of observation.}
\pagebreak
\item{On shore station:\\
      Issues such as:
      \begin{itemize}
      \item lab space,
      \item pier availability,
      \item ship and submarine availability,
      \end{itemize}
      as well as general support have to be considered.\\
      The cost of access to these supports is site dependent and has
      to be taken into account in the cost evaluation of the project.}
\end{enumerate}


\subsubsection{Test site}

To test and implement the feasibility studies, the environment
of a site is even more critical. We have started to use a site in the Toulon
area (off-shore from
La Seyne) at a depth around 2400~m (see 42$^\circ$50'N-6$^\circ$10'E in
fig.~\ref{carte-toulon}).
This site has several environmental advantages:
\begin{itemize}

\item IFREMER has a laboratory at La Seyne with piers, ships and submarines. A
civilian ship (supply-type with dynamical positioning) is also available. 

\item France T\'el\'ecom C\^able ship is based there.

\item INSU-CNRS have three ships which can be used in this area.
\end{itemize}

\subsection{Mechanical handling of the detector: deployment, recovery and
positioning}

\subsubsection{General remarks}

The detector is an array of optical modules deployed close to
the sea bottom. The physics requirement (muon threshold, up/down discrimination,
calorimetry mea\-su\-re\-ment...) will constrain the geometry of
the array. Different geometries such as strings and more elaborate structures
are shown in figure~\ref{trames}. 
We will have to deploy a
substructure of this detector alone and with an electro-optical cable connected
to it. 
One must also be able to make electrical interconnection of the many
substructures and be able to recover part of the detector for
servicing. A large scale detector with one 
electro-optical cable per substructure is unrealistic in terms of cost, 
deployment and recovery. However, a few
electro-optical cables for the whole detector are required to handle the rate
and to insure redundancy.

We will focus, in the present phase of the project, on a simple string-like
substructure of optical modules
a few hundred meter high equipped with a few tens of photo-multipliers. More
elaborate substructures will be thought of for the future.

As we intend to keep the detector under water for several years, a special
attention to corrosion problems will be given with the help of our IFREMER
collaborators and quality assurance procedures will have to be set.

The detector as a whole should be aligned and the relative position of the
different optical modules has to be monitored with an accuracy of 20~cm
(see \ref{Pos-opt-modules}).

\subsubsection{Step one: deployment and recovery of an elementary substructure}
                             
We will work on a simple substructure made of one string equipped with 
20-30~optical
modules and all the cables and the containers needed for the normal 
operation of the optical modules. We intend to use a string having
all the complexity of a final substructure even if all the photo-multipliers and
all their associated electronics might not be installed.

Vertically erected by its own buoyancy, the string is anchored on the sea
bottom by a suitably dimensioned ballast. A schematic drawing is shown in
figure~\ref{fig9jjd}. Apart from the optical modules, the string includes
also current-meters, tilt-meters, accelerometers and compasses
to learn about its
dynamical behavior during deployment, operation and recovery.

The anchoring system should also keep in position the electro-optical cable
connected to the shore, which delivers the necessary power and ensures
the data transmission through optical fibres. It has to be designed in such 
a way 
that recovering and re-deployment of the structure is feasible.

Deployment and recovery in the deep sea are difficult operations. The complexity is
increased by the electro-optical cable handling. It should be in principle 
possible to
extrapolate the procedure routinely used by scientists deploying acoustic
detectors at depth.

We consider this phase as a major step toward the km-scale detector. This step
includes setting up a shore station (at Les Sablettes at La Seyne using the
France T\'el\'ecom C\^ables terminal) and installation of acoustic beacons  
for the positioning of the string.


\subsubsection{Step two: installation of a three dimensional array}

A possible set-up of the demonstrator is shown in figure \ref{fig10jjd}.
Whatever a final choice for the substructure may be, one is faced with the
under-water interconnection of different substructures. Indeed, schemes
considering one
electro-optical cable per substructure or all the
interconnections made before a deployment of all the substructures at once, are
unrealistic and cannot be extrapolated to a large scale detector. 


In this step, one has to test the submarine connection. From the existing 
know-how,
it is much easier to think in terms of electric connection than in
terms of optical connection.

Some electric submarine connectors exist. We will have to check if they
match our
needs, in particular in terms of lifetime. Other connectors can be considered
using an electro-magnetic coupling but they still have to be developed.

Following the advice of IFREMER experts, we will use
a manned submarine ({\it e.g.} the Cyana which can be operated down to 3000~m)
to start with,
and once experience is gained,
the submarine will be replaced by a remotely operated vehicle (ROV) which can
be operated down to 6000 m.

The laying out of an interconnecting deep sea cable has to be carefully 
prepared. Preliminary work will start in
1997 and operational tests are expected in 1998.

Once one knows how to install substructures and how
to interconnect them, one should be able to build a large scale detector.

\subsubsection{Positioning of the optical modules}
               \label{Pos-opt-modules}

We need to know the relative position of the optical modules within 20~cm.
This corresponds to the precision one can expect for the relative timing of
photo-multipliers with a few photoelectrons. The angular resolution 
of the detector depends on the timing accuracy.

This can be achieved in principle with a triangular sonar base and acoustic
detectors. Such a system is under study.
\begin{itemize}
\item{The time resolution is proportional to the sonar frequency, 
      so a high
      frequency is favo\-rable. However, the sound wave attenuation length is 
      proportional to the second
      power of the frequency and the power to be supplied increases also with
      frequency. A solution has to be defined and tested.}
\item{The geometrical implementation of sonars and hydrophones,
      as well as the number of hydrophones needed for a given
      structure, should be
      optimized . The help of tilt-meters and compasses may decrease the
      number of hydrophones required well below one per optical module.}
\end{itemize}
We plan to take advantage of the know-how of different groups working in the 
field. There may be alternative solutions to internal alignment. For example, a
pulsed light source could in principle achieve the same goal. It remains to be
proven and compared. One also has to get an absolute positioning of the whole
detector to be able to point to individual sources. Solutions exist
with acoustic devices connected to surface detectors coupled to 
differential GPS.

\subsubsection{Mechanical studies of substructures}

We have listed some of the problems which have to be taken care of in the
construction of the substructures (corrosion, current effects,
deployment\ldots). We will make simulations and tests with mock-up and existing
strings to understand the behavior of detectors and structures exposed
to the deep sea current and to local sea conditions
when the structure is deployed. We have already acquired experience
from the first series of tests we have performed.
\subsection{Optical modules}
An optical module consists of a pressure-resistant glass sphere housing 
a photo-multiplier embedded in silicon gel to ensure a good optical coupling.
A high permittivity alloy cage surrounds the tube, shielding it against 
the Earth magnetic field (see fig.~\ref{modulopt}). A DC/DC converter
supplying power to the PMT is also included in the sphere.
Signal outputs, HV monitoring, etc. are sent through
water and pressure-resistant connectors to the outside world. A later version
using digital read-out electronics will also house the front-end digitizer
board described later in this document.

The choice of photo-multiplier will be based on several parameters:
\begin{itemize}
\item{the photo-cathode size, to maximize sensitivity;}
\item{the anode pulse shape, which has an impact on trigger system;} 
\item{the overall quantum efficiency, as well as the response uniformity
      and the sensitivity to residual Earth magnetic field 
      (once the PMT is shielded);}
\item{the Transit Time Spread at the single photo-electron (SPE) level, which
      contributes to the event reconstruction efficiency;}
\item{the SPE pulse height and energy resolution;}
\item{the linearity and dynamic range. 
      The Cherenkov light 
      detected by an optical module will vary from a SPE level to an amount
      causing saturation of
      the anode signal, depending on the distance to the trajectory, 
      on the energy of the 
      muon and on the proximity of the hadronic shower. Output signals from
      one or several intermediate dynodes will be needed to cover the whole
      dynamic range;}
\item{the dark current level, which affects the coincidence rates 
      between PMTs;}
\item{the rate of pre- and after-pulses, which can mimic {\it e.g.} signals
      from muon bundles.}
\end{itemize}

In order to characterize optical modules we have set up various testing 
facilities.

A test bench consisting of a couple of dark boxes with and without
magnetic shields, in which the PMTs are exposed to homogeneous
illumination coming from red, green or blue LEDs or very fast solid state laser 
pulsers (to measure time characteristics).
The main PMT characteristics are measured and compared:
dark count rate and spectrum, rate and time structure of 
pre- and after-pulses, signal pulse shape, photo-multiplier gain, 
SPE resolution, Earth magnetic field effects, 
magnetic shielding
efficiency, SPE transit time spread, overall relative 
efficiency, linearity and dynamic range, \ldots. 

Another dark
box equipped with a mechanical system allowing a blue LED to scan the 
active area of the photo-cathode is used to control the homogeneity of the
parameters listed above as a function of the position of the light spot on
the photo-cathode.

A water tank is used to study the overall response of the optical
module to Cherenkov light emitted by vertical cosmic muons in water.
The module can be rotated around a horizontal 
axis, making it possible to measure the optical module response 
at different muon angles of incidence. 
This set-up has also been used in the 200 GeV M2 muon beam at CERN.

Several types of photo-tubes have been tested.
We have extensively studied the performances of optical modules  with  
Hamamatsu R2018-3 15$^{\prime\prime}$ photo-tubes. Such large area tubes
fit well in the largest available 
pressure-resistant glass spheres (17$^{\prime\prime}$). 
However they 
show important drawbacks. The signal amplitude strongly depends on 
the distance to the photo-cathode pole of the photon conversion point. These
PMTs were also shown to be sensitive to the residual Earth magnetic field.  
Some of these photo-tubes showed significant pulse shape variations
even for a constant amplitude, and the dark noise rate has shown
to be significant and very unstable in time. Moreover, the dynode structure
seems to be too
fragile to safely cope with the mechanical stress likely to occur 
during a deployment at sea.

Therefore, we have started to test existing commercial 
8$^{\prime\prime}$ PMTs
from Hamamatsu and EMI. First measurements indicate that they do not 
suffer from the same flaws as their 15$^{\prime\prime}$ counterparts. 
Hamamatsu R5912 (10 stages) and R5912-02 (14 stages) and EMI
9353KB (12 stages) and 9355KB (14 stages) are being
tested and give satisfactory preliminary results, namely: stable behavior,
low dark noise ($<$~1kHz), good efficiency and homogeneity, good SPE
resolution ($\approx$~30\%), good SPE time resolution ($<$~3~ns~FWHM) and no 
measurable
effect of the residual Earth magnetic field.  
8$^{\prime\prime}$ PMTs are used extensively in site
measurement tests, so some experience on their behavior in-situ has
been gained.
\subsection{Data transmission, trigger and acquisition}
Dedicated electronics and data acquisition has been developed for the 
stand-alone test programme.

For the demonstrator, analog and digital data transmission schemes are 
both being studied. For the first structure that will be deployed and connected 
to an electro-optical cable, we plan to use an analog transmission. Its 
intrinsic robustness makes it easy to implement and suited to our short term 
needs. However, we have reached the conclusion that only a digital data 
transmission can meet our requirements for a km-scale detector.

\subsubsection{Electronics and acquisition system of the stand-alone tests}

The electronics and acquisition system required for our stand-alone 
tests have been 
developed around a MBX~9000 acquisition board from MII, equipped with 
a 16~MHz MC68306 processor from Motorola, one~Mbyte of PROM and up 
to two~Mbytes of RAM
for data storage. Digital and analog I/O's including two serial links are 
available for dialogue with each specific equipment such as current-meter,
acoustic modem and the extension board that holds the electronics needed for 
each test.
A Unix-like real-time operating system is running on the processor. The
acquisition is written in C. 
A configuration file describing the test sequence, is read by the acquisition
program at the startup of the processor.

\subsubsection{The electro-optical cable}

We have the opportunity to use an already existing cable that is equipped with
four mono-mode fibres. The measured attenuation is 
0.33~dB/km at $\lambda$ = 1310~nm. The required cable length to reach
the test site is around 40~km.
 
\subsubsection{Analog link}

Up to now, we have studied an analog transmission
scheme using  direct modulation of distributed feedback
(DFB) laser diodes with frequency modulation of several carriers which are
multiplexed (FDM). We tested two systems,
a 3-channel multiplexing prototype
built by IDREL, coupled to a wide-band ORTEL optical link
and a THOMSON 4-channel prototype.

 The 3-channel prototype by IDREL, with carriers at 1.35, 1.50 and 1.65~GHz 
was tested and improved. With a 35 km  fibre link (0.36dB/km), the
linearity was measured and the dynamic range is $\simeq$ 30. The time
dispersion is  $\leq$ 1.3 ns (RMS). If we use the available cable of
40 km, with a stronger attenuation (0.5dB/km) the dynamic range would be
strongly diminished ($\simeq$ 3).

The THOMSON prototype TER7000 transmits on four carriers, between 300 and
900~MHz, and four analog channels with 50~MHz BW each.
The measured performance with a
35 km fibre is a dynamic range of 40 to 50 and a time dispersion of
1.3 ns (RMS).\\

Using the available cable we are considering the transmission of 
eight PMT signals using direct modulation and a 
simple method of mixing, without RF multiplexing.\\

Increasing the number of frequency-multiplexed channels
appears to be non trivial. To date, a digital solution seems to be the only way
to meet the requirements of a km-scale detector.

\subsubsection{Digital link}

A possible digital architecture for a large scale detector could be organized
as a tree-structured network of MCMs (Main Control Modules), SCMs (String
Control Modules), LCMs (Local Control Modules) and DOMs (Digital Optical
Modules) (see e.g. the LBNL-JPL group proposal~\cite{ATWR}). The MCMs would be
connected to the shore via electro-optical cables. Interconnections between
control modules at the same level in the network hierarchy would ensure a path
redundancy for data/control information.

A possible approach to digitize the signals at the optical module level is
to have LCMs connected to several DOMs via bidirectional links. Each DOM
produces a local trigger which is transmitted to
  the LCM. A higher level trigger produced in the LCM is
  sent back to the DOM's in order to enable digitization and 
  data transmission.

\begin{itemize}
\item {{\bf The Digital Optical Module (DOM)}:\\
In \cite{ATWR} it is
proposed to develop a DOM based on a
ASIC designed at LBL, the Analog Transient Waveform Recorder (ATWR).

We are developing a similar architecture, in a circuit called
 Analog Ring Sampler (ARS). The ARS ASIC consist of
 an array of 128 capacitor cells which samples and memorizes
analog input signals. The sampling
frequency is adjustable from 300~MHz to 1~GHz. 
The ARS has five channels of 128 cells, one channel for the signal
coming from the PMT anode, two channels for the signals coming from two 
intermediate dynodes, one channel dedicated to time stamping in which a 
stable 20 MHz clock is recorded and the last channel being used for accurate 
pedestal subtraction.

In acquisition mode, the ARS is constantly sampling the PMT signals
as a ring memory. When the PMT signal crosses a comparator
threshold corresponding to a fraction of SPE the ARS stops overwriting its
cells. The comparator signal is sent to the Local Control Module (LMC)
where the trigger is built.

In order to reduce the amount of data to be transmitted, 
pulse shape discrimination (PSD) is performed in parallel to trigger building.
The decision is made whether to provide only time and 
charge (SPE Mode) or to fully
digitize pulse shapes departing from SPE (Waveform Mode).

The shape discrimination consists of a threshold comparator, a time-over-
threshold comparator and a multi-pulse detector in order to recognize three 
kinds of shapes: high pulses, long pulses and multiple pulses.

The motherboard, inside the DOM, supports two channels (ARS) in ``flip-flop''
operation, to reduce dead-time. It also drives the
power supplies and surveys the current, voltage and temperature,
generating an alarm signal if necessary.
It receives the clock from the LCM and supports the slow control.}\\

\item {{\bf The Local Control Module (LCM)}:\\
The LCM is the next node in the digital network of the detector array. It 
is connected to about 10~DOMs on one hand, to the next digital node (e.g. 
String Control Module) on the other hand. It takes care of power distribution,
slow control commands and data, trigger building and formatting and 
transmission of DOM data to the SCM. 

Any DOM which is locally triggered sends a coincidence request to the LCM.
The LCM looks for a time coincidence with a request from any of 
the other DOMs connected to it and sends back a data request signal to the DOMs 
involved. The data are then digitized and sent to the LCM. A scheme where 
the DOMs are grouped in close clusters (e.g. pairs or quadruples) allows 
tight time coincidences. The accidental coincidence rate coming from $^{40}$K 
can be kept around 100~Hz and the system can cope with up to a few 100~kHz
bioluminescence bursts before suffering from significant dead time.

The front end PSD ensures that the data flow will be minimal. In SPE mode, an 
event represents about 30~bytes of data. In Waveform mode (e.g. 1$\%$ of the 
time), it goes up to 600~bytes. The average data flow will not exceed 
50~kbyte/s per LCM. This allows the use of Ethernet(-like) protocols for the 
detector network.}
\end{itemize}

\subsection{Slow Control}

Optical modules (PMTs voltages, temperatures), calibration, positioning 
systems and sensors
for detector geometry together with  sea parameters are controlled and read out
by the slow control system which gives a 
user in the shore station all the needed graphical and numerical information to
monitor and control the detector.

The slow control system will be based on an industrial field-bus network 
technology
providing reliability, robustness and scalability features to such an embedded
detector control. The 
bandwidth of these field-buses will fit our future system extensions 
(sensors, control devices).
The link between the undersea part of slow control and the shore station
will be done via the electro-optical cable.
An interface using an acoustic modem will provide the user means for a 
stand-alone (without connection to shore) pre-diagnosis of the slow control
system.

\subsubsection{Slow Control network}

To transfer data from the sensors and control-command messages to the optical
modules (OM), we will use an industrial field-bus network techno\-logy called
WorldFip designed in the early 90's to cope with sensors and actuators
control in a factory environment. The main features of WorldFip are~: 

\begin{itemize}
\item Deterministic field-bus, the communication protocol used in
WorldFip allows a mix of periodic and aperiodic data transfers. In the
first case, the time schedule of periodic variables is warranted.  

\item High speed sensor/actuator communication (32~kb/s, 1~Mb/s, 2~Mb/s) 
 
\item Long distance connection between WorldFip nodes (up to 1~km at 1~Mb/s)

\item Redundancy, WorldFip support bi-medium connection on two different
shielded copper twisted pair. A switch to the best medium is
automatically done by the WorldFip protocol chipset.  

\item Reliable bus arbitration policy based on multiple bus arbiter
  modules on the same WorldFip network.  
\end{itemize}

We increase the robustness of the slow control architecture by
segmenting the network by using an active network star repeater. In
this architecture, each LCM container of the string is directly
connected to the main Slow Control data acquisition system.  A test
bench for WorldFip network implementation evaluation is already setup
using two PC's and WorldFip to PC interfaces from CEGELEC (FullFip and
MicroFip).

\subsubsection{Slow Control bridge system}

The Slow Control sensors (attitude and environment sensors) will
mainly be read through serial interfaces (RS232 and RS485) while the
Optical Modules (OM) will be locally controlled by a micro-controller (for
temperature, DC, calibration LED) communicating through RS485
differential lines.  The interface between these serial lines and the 
 WorldFip network is performed by using a MBX9000-40 acquisition board
(see paragraph 3.4.1) which can support 8 serial lines and is interfaced to
WorldFip network (add-on module). One serial line will be dedicated to the
string geometry sensors readout (tilt-meters, magnetic field..) and the others
will be individually connected to the optical modules. One
bridge system is included in each LCM container of the string.

\subsubsection{Slow Control acquisition system}

The slow control data acquisition (DAQ) system is build around a G96
module with a Motorola 68k processor running an embedded real-time OS9
operating system.  The Gespac G96 bus connects this processor module
to a G96 bi-medium WorldFip module (linked to the WorldFip active star
repeater) and the I/O modules controlling energy distribution. This
DAQ system is located in the Main Electronic Container located at the
bottom of the string.  An electro-optical cable connects this embedded
Slow Control DAQ system to the shore station which receive the
Slow Control data and provides a user interface for the Slow Control
system.

\newpage
\section{Simulation and detector optimization} 
\subsection{General remarks}

A Monte Carlo simulation of our detection system is necessary to give us a way
to study how to:
\begin{itemize}
  \item optimize the geometry of the detector,
  \item adjust trigger schemes,
  \item tune track reconstruction algorithms,
  \item estimate energy resolution,
  \item investigate effects of backgrounds:
    \begin{itemize}
    \item atmospheric $\nu$'s and $\mu$'s,
    \item $^{40}$K and bioluminescence,
    \end{itemize}
  \item estimate the detector sensitivity to various neutrino sources.
\end{itemize}
The Monte Carlo program has to deal with Cherenkov photons generated by 
neutrino
induced muons together with the electromagnetic and hadronic showers they
caused while traveling through the detector and its neighborhood. The
program
simulates the propagation of these photons in water and the signal they induce
on the 3-D matrix of optical modules the detector is made of. 
Because the
number of secondary particles accompanying the primary muons increases
dramatically with energy, a full simulation {\em \`a la} GEANT (which needs
anyway to be modified to be reliable above 10 TeV) becomes quickly
CPU-time prohibitive
as energies of order a few~TeV are reached (see figure~\ref{muons}). As large 
samples of events are
needed in order to study resolutions and efficiencies, two approaches are
currently being investigated:
\begin{itemize}
  \item a parameterization of the Cherenkov light distributions of 
        electromagnetic
        and hadronic showers, as pioneered by SiEGMuND (Baikal
        collaboration)~\cite{wiebusch},
  \item generated Cherenkov light look-up tables, 
        as used in RAVEN (AMANDA)~\cite{moorhead-raven} and 
        the KM$^3$ simulation 
        programs~\cite{moorhead-km3}.
\end{itemize}
%

%
\subsection{Geometry}

Variations of geometrical parameters like the number of PMTs, their
density, their spatial distribution, their size and type, the distance between
groups of PMT, are being studied in order to optimize 
trigger and reconstruction efficiencies as well as angular
and energy resolution. The detector effective area A$_{eff}$ is defined as
A$_{eff} = f \cdot S$, with $S$ the surface
inside which muons are generated uniformly and $f$ the fraction of
events that pass the trigger and/or give a successful track reconstruction.
It depends strongly on the
requested angular reconstruction accuracy and increases substantially with
energy. 
\subsection{Trigger studies}

Several trigger configurations will be necessary to cover different physics
channels of interest. For the demonstrator, the trigger will be an easy 
problem to solve. For a large scale detector, the main emphasis at 
the beginning will be put on 
neutrino induced muons with energy deposition all along their
trajectories. Other more difficult channels as
supernovae neutrino bursts and 
``double'' bang structures for $\nu_\tau + N \rightarrow \tau + X$
events will be studied later.
Most of these channels satisfy the simple criterion of a minimal number of
photo-multipliers (PMT) above a threshold set below 1 photo-electron (PE). The
goal is to trigger efficiently on muons in the largest possible volume for a
given cost of the detector.\\
The main optical backgrounds (bioluminescence and natural radioactivity) are
uncorrelated and mostly
contribute to 1 PE signals, so the trigger can be designed with tight
coincidences and/or a signal threshold greater than 1 PE. We need to
quantify the trigger efficiency for a minimal number of hit PMTs estimating the
detector effective area by computations based on Monte Carlo
simulations. Several PMT configurations are currently under study.
\subsection{Muon track parameters}
\begin{itemize}
\item{{\bf Muon track reconstruction}:\\
The muon which triggered the system has to be reconstructible ({\it i.e.} we 
need
sufficient information to efficiently determine the muon direction with a good
enough angular accuracy). The effective area used to evaluate the different
geometric configurations has to be corrected for reconstruction efficiencies.\\
Several reconstruction algorithms, all based on the
characteristics of Cherenkov light emission, are currently under study. A muon
track is characterized 
by five parameters (one space point and two angles), so at least five
hit PMTs are needed. For a given track, 
the arrival
time of the Cherenkov light on PMT $i$ can be calculated as:
\[
  c~t_i = L_i + d_i \tan\theta_c
\]
where $d_i$ is the impact parameter of the muon track w.r.t. the PMT, $L_i$ the
distance between the track point at a distance $d_i$ from the PMT and the track
point corresponding to the triggering time (t=0), $c$ the speed of light in
vacuum and $\theta_c$ the Cherenkov angle (see fig.~\ref{fig:recon}).
The problem then boils down to a minimization of:
\[
  \chi^2 = \sum_{i} \frac{1}{\sigma_t^2}\left(t_i^m-t_i\right)^2
\]
with $t_i^m$ the measured arrival time of Cherenkov light on PMT $i$ and
$\sigma_t$ the time resolution of the PMT ($\sigma_t = \sigma_{1PE}/\sqrt{N_i}$
with $\sigma_{1PE}$ the PMT time resolution for a 1 PE signal and $N_i$ the
signal charge measured in units of PEs).}
\item{{\bf Muon energy estimate}:\\
In the above expression, the charge of the PMT signals enter only indirectly
through the time resolution. Actually, this extra information is complementary
though it is more subject to big fluctuations. We reserve its use to select PMTs
which are to be used in the fit and also to estimate the muon energy. A way to
find the most probable energy of the muon could be to use a maximum likelihood
method {\it \`a la} Baikal. They define a function:
\[
  L(\log E_\mu) = \prod_{i=1}^{N} W_i(E_\mu,A_i,R_i,\theta_i,\phi_i)
\]
with $W_i$ the probability to observe an amplitude $A_i$ on PMT $i$ located at a
distance $R_i$ from the muon trajectory and with angles $\theta_i$ and $\phi_i$
w.r.t. it if the energy of the muon is assumed to be $E_\mu$; N is the overall
number of PMTs at a distance $R < R_{min}$ from the muon trajectory.}
\item{{\bf Pending issues to be studied}:
\begin{itemize}
  \item the impact of non-correlated backgrounds on the reconstruction
        efficiency and trigger rates, and the optimization of selection and
        reconstruction algorithms in order to minimize the effects of such
        backgrounds,
  \item the reconstruction efficiency with emphasis on reconstruction errors.
        The issue here is to succeed in rejecting single downward going 
        atmospheric
        muons as well as multiple downward going muons which, if too close in 
        time,
        could fool the reconstruction algorithm and be taken as a single 
        upward going
        muon. In this latter case, a shape analysis provided by a fast enough
        sampling of the PMT signals could prove to be very helpful,
  \item the reconstruction efficiency for very high energy 
        muons (PeV and above),
        for various geo\-me\-tries.
\end{itemize}
}
\end{itemize}

\newpage
\section{The ANTARES Collaboration} 
\subsection{Collaborating institutes}

The collaboration consists of groups of particle physicists and engineers from:
\begin{itemize}
  \item {CPPM Marseille (IN2P3-CNRS/Universit\'e de la M\'editerran\'ee):\\
  C.~Arpesella, E.~Aslanides, J.J.~Aubert, S.~Basa, V.~Bertin, M.~Billault,
  P.E.~Blanc, A.~Calzas, C.~Carloganu,
  J.~Carr, J.J.~Destelle, F.~Hubaut, \\
  E.~Kajfasz, R.~Le~Gac,
  A.~Le~Van~Suu, L.~Martin, C.~Meessen, F.~Montanet,
  Ch.~Olivetto, P.~Payre, R.~Potheau, M.~Raymond, M.~Talby, E.~Vigeolas.}
  \item {DAPNIA-DSM-CEA (Saclay):\\
  R.~Azoulay, R.~Berthier, F.~Blondeau, N.~de~Botton, P.H.~Carton, \\
  M.~Cribier,
  F.~Desages, G.~Dispau, F.~Feinstein, P. Galumian, L.~Gosset, 
  J.F.~Gournay, D.~Lachartre,
  P.~Lamare, J.C.~Languillat, J.Ph.~Laugier, \\
  H.~Le~Provost, D.~Loiseau,
  S.~Loucatos, P.~Magnier, J.~McNutt, P.~Mols, L.~Moscoso, P.~Perrin, 
  J.~Poinsignon, Y.~Sacquin, 
  J.P.~Schuller, J.P.~Soirat, A.~Tabary, D.~Vignaud, D.~Vilanova.}
  \item {Instituto de F\'{\i}sica Corpuscular -- CSIC-Universitat de
  Val\`encia:\\
  R.~Cases, J.J.~Hern\'andez, S.~Navas, J.~Velasco, J.~Z\'u\~niga.}
  \item{Oxford University -- Physics Department:\\
  D.~Bailey, S.~Biller, B.~Brooks, N.~Jelley, M.~Moorhead, D.~Wark.}
\end{itemize}
of groups of sea scientists from:
\begin{itemize}
  \item {Centre d'Oc\'eanologie de Marseille:\\
  F. Blanc, J.L. Fuda, L. Laubier, C. Millot.}
  \item {IFREMER:\\
  J.F.~Drogou, D.~Festy, G.~Herrouin, L.~Lemoine, F.~Mazeas, P.~Valdy.}
\end{itemize}
of groups of astronomers and astrophysicists from:
\begin{itemize}
  \item {IGRAP (INSU-CNRS):\\
  Ph.~Amram, J.~Boulesteix, M.~Marcelin, A.~Mazure, R.~Triay.}
  \item {DAPNIA-DSM-CEA (Saclay):\\
  Ph.~Goret.}
\end{itemize}

The collaboration has also the support of other organizations as France
T\'el\'ecom C\^ables, CSTN, CTME.

Geophysicists have shown an interest to share some technological developments
(IPG - Paris, Sophia Antipolis - Nice).

\subsection{Schedule}

The existing programme has been approved for the
demonstrator implementation and will be performed over 1997, 1998 and 1999.

We would like to complete in the first year the apparatus dedicated to the
site tests, then in the second year to have one string with an electro-optical
cable and in the third year to complete the 3-dimensional demonstrator. In the
same time, one should develop the tools for the next phase.

\subsection{Enlargement of the Collaboration}

New collaborators are needed, as soon as possible, to
bring new ideas and to participate in these developments. There are lots of
tasks which have to be achieved as in a standard particle physics experiment.

The collaboration is actually sharing the most urgent tasks and it has been
decided to redistribute every year the responsibilities, to be able to take
into account the changes in the collaboration.

After the completion of the demonstrator, we would like to build a km-scale
detector step by step. The next stage may be a high muon energy threshold
detector of a large effective surface with several hundred photo-multipliers.
This stage can be achieved with new collaborators on a reasonable time scale.

\newpage
{\bf \LARGE APPENDIX}\\
\appendix
\addcontentsline{toc}{section}{\numberline{} APPENDIX - Neutrino-matter
interaction}
{\bf \Large Neutrino-matter interaction}
  \begin{itemize}
  \item {The inclusive charged current cross-section for 
  $\nu_\mu + N \rightarrow \mu^- + X$ is given by \cite{quigg}:
  \[
     \frac{d^2\sigma_{\nu N}}{dxdy} = \frac{2G^2_Fm_NE_\nu}{\pi} 
     \frac{M^4_W}{(Q^2+M^2_W)^2} \left[ xq(x,Q^2)+x(1-y)^2\bar q(x,Q^2) \right]
  \]
  where $x = Q^2/2m_N\nu$ and $y = \nu/E_\nu$ and with
    $Q^2$, the square of the momentum transfer between the neutrino and muon,
    $\nu = E_\nu - E_\mu$ the lepton energy loss in the lab frame,
    $m_N$ the nucleon mass,
    $M_W$ the $W$-boson mass
    and $G_F$, the Fermi constant.
  For an isoscalar nucleon, in terms of valence ($v$) and sea ($s$) parton
  distribution functions:
  \[
    q(x,Q^2) = \frac{u_v+d_v}{2} + \frac{u_s+d_s}{2} + s_s + b_s
  \]
  \[
    \bar q(x,Q^2) = \frac{u_s+d_s}{2} + c_s + t_s
  \]
  The energy dependence of the cross-section is shown in figure
  \ref{fig:quigg-4} (see ref. \cite{quigg}).\\

%
%

  At low energies, the cross-section grows linearly with the neutrino energy:
  \[
    \sigma_{\nu N}(E_\nu) \simeq 0.67~ 10^{-38} E_\nu[GeV]~cm^2 
  \]
  \[
    \sigma_{\bar\nu N}(E_\nu) \simeq 0.34~ 10^{-38} E_\nu[GeV]~cm^2    
  \]
  At energies such that $E_\nu \gg M^2_W/2m_N \approx 5$ TeV,
  the $W$ propagator limits the
  growth of $Q$ to $<Q^2> \sim M^2_W$ and makes the cross-section to be
  dominated by the behavior of distribution functions at 
  small $x$ ($x \sim M^2_W/2m_NE_\nu<y>$).\\
  The H1 and ZEUS collaborations at HERA measured the proton
  structure function $F_2(x,Q^2)$ via charged current $e-p$ scattering,
  for $Q^2$ in the range $[1.5,5000]$ GeV$^2$ with $x$ down to $3~10^{-5}$
  at $Q^2 = 1.5$ GeV$^2$ and $x$ down to $2~10^{-2}$ at 
  $Q^2 = 5000$ GeV$^2$~\cite{H1ZEUS}.
  These measurements can be translated into a neutrino-nucleon 
  interaction cross
  section at $E_\nu \simeq 50$ TeV and can also be used as a guide to
  extrapolate the parton densities beyond the measured ranges in $x$ and $Q^2$
  required at even higher neutrino energies. Figure \ref{fig:quigg-5}
  shows the behavior of the interaction cross section of a
  $\nu_\mu$ with an isoscalar nucleon for different sets of parton distribution
  functions. At very high energy, the newly calculated cross section (see fig.
  \ref{fig:quigg-4}) with CTEQ3-DIS \cite{cteq} 
  parton distribution functions, is more than
  a factor of 2 larger than the pre-HERA estimate calculated with EHLQ
  parton distribution functions.}
  \item {As the calculated cross section was underestimated at high energy, so 
  was the attenuation of the neutrino flux in the Earth.
  Figure \ref{fig:quigg-11} shows the interaction length 
  $\cal L_{int}$ of $\nu_\mu$'s on nucleon targets as a function of $E_\nu$.
  $\cal L_{int}$ is given by $\cal L^{-1}_{int} = N_A\sigma_{\nu N}(E_\nu)$, 
  where $N_A$ is the Avogadro number. At energies 
  greater than $\simeq$ 40 TeV, the
  interaction length becomes smaller than the Earth diameter.\\

%
%
%

  The attenuation of neutrinos in the Earth can be described by a shadow factor
  $S(E_\nu)$ given by:
  \[
    \frac{dS(E_\nu)}{d\Omega} = \frac{1}{2\pi}\exp\left[\frac{-d(\theta_z)}
    {\cal L_{int}(E_\nu)}\right]
  \]
  where $d(\theta_z)$ is the column depth of the Earth at zenith angle
  $\theta_z$.}
  \item {The energy loss of muons in matter is due to quasi-continuous (with small
  fluctuations) ionization processes but also to radiative processes
  (bremsstrahlung, pair production and photo-production) subject to
  big fluctuations (for a thorough treatment of the propagation of multi-TeV
  muons in matter see ref. \cite{lipari}). The average loss can be expressed
  as:
  \[
    \left<\frac{dE(E_\mu)}{dX}\right> = \alpha+\beta E_\mu
  \]
  where $\alpha$, representing the ionization contribution, is
  logarithmically
  increasing with energy. $\beta$ is the sum of the fractional energy radiation
  losses. For $E_\mu$ greater than the critical energy 
  $\varepsilon = \alpha/\beta$, the radiation losses dominate. Bremsstrahlung and
  pair creation contributions to $\beta$ reach an asymptotic value at high
  energy, but it is not so for the photo-nuclear contribution which
  keeps rising with energy.\\
  If $\alpha$ and $\beta$ are taken to be energy independent ($\alpha \simeq 
  2.0~10^{-3}$ GeV~cmwe$^{-1}$, $\beta$ = 3.9~10$^{-6}$ cmwe$^{-1}$ in 
  rock so $\varepsilon \simeq 500$ GeV), then 
  the range of the average loss for a muon of initial energy $E_\mu$ and final
  energy $E^{min}_\mu$ is given by:
  \[
    R_{<\Delta E>}(E_\mu;E^{min}_\mu) = 
    \int_{E^{min}_\mu}^{E_\mu}\frac{dE_\mu}{\left<\frac{dE(E_\mu)}{dX}\right>}
    \simeq \frac{1}{\beta} \ln \left[\frac{\alpha+\beta E_\mu}
    {\alpha+\beta E^{min}_\mu}\right]
  \]
  For $E_\mu \ll \varepsilon$, the range of muons is correctly reproduced by
  $R_{<\Delta E>}(E_\mu) \propto E_\mu$ as ionization processes dominate.\\
  For $E_\mu \gg \varepsilon$, $R_{<\Delta E>}(E_\mu) \propto \ln(E_\mu)$.
  However,
  in this regime, the radiation losses dominate and because of their 
  big fluctuations:
  \begin{itemize}
     \item the average range $<R(E_\mu;E^{min}_\mu)>$ of the
      muons is smaller than the range of the average loss 
      $R_{<\Delta E>}(E_\mu;E^{min}_\mu)$,
     \item the width of the range distribution increases with energy,
  \end{itemize}
  so the range of the average loss only poorly reproduces the actual range
  of high energy muons.
  In order to estimate the range of high energy neutrinos correctly, it is
  necessary to use a Monte-Carlo method to propagate them.}
  \end{itemize}
  Two approaches can be followed to calculate muon rates $N_\mu$ in a detector:
  \begin{itemize}
    \item Analytically,
     the probability for a neutrino of energy $E_\nu$ to give a muon of energy
     $E_\mu > E^{min}_\mu$ in the detector is given by:
     \[
       P_\mu(E_\nu;E^{min}_\mu) = N_A \sigma^{CC}_{\nu N}(E_\nu) 
       <R(E_\mu;E^{min}_\mu)>
     \]
     where $<R>$ is the average muon range in rock.\\
     The rate $N_\mu$ of muons induced by an isotropic neutrino flux 
     $\frac{dN_\nu}{dE_\nu}$ in a detector of effective area $A$ is given by:
     \[
      N_\mu = A\int d\Omega dE_\nu P_\mu(E_\nu;E^{min}_\mu) S(E_\nu) 
      \frac{dN_\nu}{dE_\nu} 
     \]
     where the shadow factor is integrated over the relevant solid angle.
     Figure \ref{fig:quigg-23} shows how the product 
     $P_\mu(E_\nu;E^{min}_\mu)S(E_\nu)$ evolves as a function of $E_\nu$ for
     two muon threshold energies $E^{min}_\mu$ of 1 and 10 TeV respectively.
%
%
    \item Using
     a Monte Carlo program in which a
    given neutrino flux is made to interact with the Earth, the induced muons
    are then propagated to the detector. 
    It fully takes into account the large fluctuations occuring at high
    energy. This method was prefered to the analytical one to compute
    the rate results presented in section \ref{sec:rates}. 
  \end{itemize}
\newpage
\addcontentsline{toc}{section}{\numberline{} References}

\newpage
\begin{table}[p]
  \begin{center}
    \begin{tabular}{|r|c|c|}
      \hline
       & & \\
      Energy & $\Phi_0$ & Atm. neutrinos \\
      threshold  & (cm$^{-2}$\,s$^{-1}$) &
                            (cm$^{-2}$\,s$^{-1}$\,sr$^{-1}$)  \\
        & & \\
      \hline
        & & \\
      10\,GeV & $1.3 \cdot 10^{-13}$ & $1.5 \cdot 10^{-13}$ \\
      100\,GeV & $1.2 \cdot 10^{-13}$ & $5.3 \cdot 10^{-14}$ \\
      1\,TeV & $6.6 \cdot 10^{-14}$ & $5.9 \cdot 10^{-15}$ \\
      10\,TeV & $1.5 \cdot 10^{-14}$ & $1.4 \cdot 10^{-16}$ \\
       & & \\
      \hline
    \end{tabular}
  \end{center}
  \caption{Values of the neutrino induced muon flux $\Phi_0$ (see text) for 
  different energy
  thresholds and for a differential spectral
  index $\gamma = 2.2$ together with the
  flux of muons induced by atmospheric neutrinos
  averaged over the whole $2 \pi$ downward hemisphere for the same thresholds.}
  \label{lmtb:tab1}
\end{table}
\begin{table}[p]
\begin{center}
\begin{tabular}{|r|c|}
\hline
 &  \\
Energy &  ${\cal L_p}$ \\
threshold &  (erg\,s$^{-1}$) \\
 &  \\
\hline
 &  \\
10\,GeV &   8.3\,10$^{34}$  \\
100\,GeV &  5.3\,10$^{34}$  \\
1\,TeV &    4.8\,10$^{34}$  \\
10\,TeV &   2.1\,10$^{35}$  \\
  &  \\
\hline
\end{tabular}
\end{center}
\caption{Luminosities of a neutrino source with a differential spectral index 
$\gamma = 2.2$ located at a distance of 1\,kpc that can be
detected at a 5 $\sigma$ level (or 10 events per year) with
an effective exposure of 1\,km$^2$\,year 
for different muon energy thresholds.}
\label{lmtb:tab2}
\end{table}
\begin{table}[p]
\begin{center}
\begin{tabular}{|l|r|c|r|c|c|}
\hline
& & & & & \\
\multicolumn{1}{|c|}{Source} & Declination & Distance & \multicolumn{1}{c|}{${\epsilon}_t$}
& Expected & ${\cal L}_p$  \\
\multicolumn{1}{|c|}{name} & \multicolumn{1}{c|}{$\delta$} & \multicolumn{1}{c|}{(kpc)} & & background
& ${\gamma}=2.2$ \\
& & & & & (erg/s) \\
& & & & & \\
\hline
& & & & & \\
{\underline{Pulsars}} & & & & & \\
& & & & & \\
Crab & 21$^{\circ}$ & 2 & 37.5\% & 0.4 & 5.1\,10$^{35}$ \\
Vela & -45$^{\circ}$ & 0.5 & 100\% & 3.0 & 1.2\,10$^{34}$ \\
PSR\,1937+21 & 21$^{\circ}$ & 5  & 37.5\% & 0.4 & 3.2\,10$^{36}$ \\
PSR\,1953+29 & 29$^{\circ}$ & 3.5 & 31.3\% & 0.3 & 1.9\,10$^{36}$ \\
PSR\,1822-09 & -9$^{\circ}$ & 0.6 & 55.1\% & 0.8 & 3.1\,10$^{34}$ \\
PSR\,1801-23 & -23$^{\circ}$ & 2.7 & 64.0\% & 1.2 & 5.5\,10$^{35}$  \\
& & & & & \\
{\underline{Binary stars}}& & & & & \\
& & & & & \\
Cyg-X3 & 41$^{\circ}$ & 10-12.5 & 16.5\% & 0.1 & (2.9-4.5)\,10$^{37}$  \\
Her-X1 & 35$^{\circ}$ & 5 & 25.3\% & 0.3 & 4.7\,10$^{36}$  \\
Cyg-X1 & 35$^{\circ}$ & 2.5 & 25.3\% & 0.3 & 1.2\,10$^{36}$ \\
SS433 & 5$^{\circ}$ & 5 & 47.2\% & 0.7 & 2.5\,10$^{36}$  \\
Vela-X1 & -40$^{\circ}$ & 1.4 & 81.7\% & 2.7 & 1.1\,10$^{35}$ \\
& & & & & \\
{\underline{SN remnants}}& & & & & \\
& & & & & \\
Crab & 22$^{\circ}$ & 2 & 37.5\% & 0.4 &  5.1\,10$^{35}$ \\
1987A & -69$^{\circ}$ & 50 & 100\% & 2.6 & 1.2\,10$^{38}$  \\
& & & & & \\
\hline
\end{tabular}
\end{center}
\caption{Detectable proton luminosities for several known galactic sources. The detector
latitude is $45^{\circ}$. ${\epsilon}_t$ depends on the declination $\delta$ of
the source.}
\label{lmtb:tab3}
\end{table}
\begin{table}[p]
\begin{center}
\begin{tabular}{|l|c|c|c|c|c|c|c|}
\hline
 & & & & & & &\\
\multicolumn{1}{|c|}{Model} & $\cos\theta_z$ & E$^{min}_\mu$ = 1\,TeV & 10\,TeV &
 100\,TeV & 1\,PeV & 10\,PeV & 100\,PeV\\
 & range & & & & & &\\
 & & & & & & &\\
\hline
\hline
 & & & & & & &\\
{\underline{Atmospheric}}& & & & & & &\\
 & & & & & & &\\
ATM~\cite{volkova}  & $[-.25,1.]$ & 16000  & 420 & 5.9 & 0.08 & 0.001 & - \\
                    & ~$[0.,1.]$~ & 12000  & 280 & 3.6 & 0.04 & 0.004 & -  \\
 & & & & & & &\\
\hline
 & & & & & & &\\
{\underline{AGNs}}& & & & & & &\\
 & & & & & & &\\
SDSS~\cite{stecker}  & $[-.25,1.]$ & 6700  & 4500 & 2000 & 400 & 39 & 1.9  \\
                    & ~$[0.,1.]$~ & 4200  & 2700 & 1100 & 180 & 12 & 0.2 \\
NMB~\cite{nellen}  & $[-.25,1.]$ & 8200 & 2300 & 190 & - & - & -  \\
                   & ~$[0.,1.]$~ & 6300 & 1700 & 130 & - & - & - \\
 & & & & & & &\\
\hline
 & & & & & & &\\
{\underline{Blazars}}& & & & & & &\\
 & & & & & & &\\
PRO~\cite{prothe-blazar} & $[-.25,1.]$ & 980 & 710 & 380 & 130 & 33 & 6.1 \\
                         & ~$[0.,1.]$~ & 510 & 360 & 180 & 48 & 7.6 & 0.5 \\
MRLA~\cite{mannheim-blazar} & $[-.25,1.]$ & 420 & 72 & 9.5 & 2.2 & 0.8 & 0.2 \\
                            & ~$[0.,1.]$~ & 330 & 52 & 5.1 & 0.7 & 0.2 & 0.02 \\
MRLB~\cite{mannheim-blazar} & $[-.25,1.]$ & 530 & 160 & 65 & 32 & 14 & 3.9 \\
                            & ~$[0.,1.]$~ & 370 & 81 & 22 & 8.8 & 2.6 & 0.4 \\
 & & & & & & &\\
\hline
 & & & & & & &\\
{\underline{Defects}}& & & & & & &\\
 & & & & & & &\\
BHSl~\cite{bhatta} & $[-.25,1.]$ & 9.9 & 7.6 & 4.6 & 2.4 & 1.2 & 0.5 \\
                      & ~$[0.,1.]$~ & 3.8 & 2.6 & 1.4 & 0.6 & 0.2 & 0.05 \\
BHSh~\cite{bhatta} & $[-.25,1.]$ & 1200 & 720 & 350 & 150 & 62 & 21 \\
                      & ~$[0.,1.]$~ & 680 & 340 & 130 & 42 & 10 & 2 \\
SIG~\cite{sigl}    & $[-.25,1.]$ & 0.16 & 0.13 & 0.08 & 0.05 & 0.035 & 0.01 \\
                   & ~$[0.,1.]$~ & 0.05 & 0.04 & 0.02 & 0.01 & 0.004 & 0.001 \\
 & & & & & & &\\
\hline
\end{tabular}
\caption{Number of neutrino induced muons per year in a 1\,km$^2$ detector
for two ranges of zenith angle $\theta_z$ and different muon energy thresholds
$E^{min}_\mu$. Diffuse neutrino sources considered here are the Earth atmosphere,
 Active Galactic Nuclei, Blazars jets and topological defects.}
\label{lmtb:tab4}
\end{center}
\end{table}
\begin{table}[p]
\begin{center}
\begin{tabular}{|lccccccc|}  \hline
Source name & RA & $\delta$ & $\gamma$ meas. & ${\epsilon}_t$ &
\multicolumn{3}{c|}{\# of expected events} \\
 & (Degrees) & (Degrees) & & & $\gamma$ meas. & $\gamma = 2$ & 
${\gamma}_{\mathrm av} = 2.12$ \\
\hline
J0204+1512 & 31.03 & 15.22 & $2.5 \pm 0.1$ &  0.412 & 0.1 & 20 & 6.1 \\
J0210-5051 & 32.69 & -50.85 & $1.7 \pm 0.1$ &  1.000 & 7700 & 170 & 52 \\ 
J0450+1122 & 72.57 & 11.38 & $2.2 \pm 0.1$ &  0.435 &  1.9 & 21 & 6.2 \\
J0531+1324 & 82.74 & 13.53 & $2.3 \pm 0.1$ &  0.423 &  2.3 & 80  &  24 \\
J1104+3812 & 166.11 & 38.21 & $1.7 \pm 0.2$ &  0.212 & 300 & 6.6   & 2.0 \\
(Mrk\,421) & & & & & & & \\
J1158+2906 & 179.67 & 29.11 & $2.0 \pm 0.5$ &  0.312 & 150 & 150  &  44 \\
J1224+2155 & 186.12 & 21.92 & $2.0 \pm 0.2$ &  0.368 & 14 & 14 & 4.2 \\
J1229+0206 & 187.25 &  2.10 & $2.5 \pm 0.2$ &  0.488 & 0 & 16   & 4.9 \\
(3C\,273) & & & & & & & \\
J1256-0546 & 194.04 & -5.79 & $2.02 \pm 0.07$ & 0.532 & 240 & 310 & 92 \\
(3C\,279) & & & & & & & \\
J1409-0742 & 212.23 & -7.88 & $2.0 \pm 0.1$ &  0.544 & 50 & 50 & 15 \\
J1513-0857 & 228.26 & -8.95 & $2.3 \pm 0.3$ &  0.550 & 0.7 & 24 & 7.1 \\
J1608+1046 & 242.17 & 10.77 & $2.4 \pm 0.3$ &  0.439 & 0.3 & 30 & 8.9 \\
J1614+3431 & 243.65 & 34.52 & $1.8 \pm 0.3$ &  0.259 & 120 & 10 & 3.0 \\
J1626-2452 & 246.57 & -24.87 & $2.3 \pm 0.2$ &  0.653 & 1.3 & 45 & 13 \\
J1635+3813 & 248.92 & 38.21 & $2.1 \pm 0.1$ &  0.212 & 10  & 32   & 9.5 \\
J1735-1312 & 263.80 & -13.21 & $2.4 \pm 0.3$ &  0.575 & 0.3 & 33 & 9.8 \\
J1911-1945 & 287.97 & -19.76 & $2.5 \pm 0.2$ &  0.617 & 0.1  & 21 & 6.2 \\
J1934-4014 & 293.67 & -40.24 & $2.4 \pm 0.2$ &  0.821 & 0.2  & 21 & 6.4 \\
J2023-0836 & 305.96 &  -8.61 & $1.5 \pm 0.2$ &  0.548 & 19000 & 27 & 8.2 \\
J2058-4657 & 314.52 & -46.96 & $2.4 \pm 0.4$ &  1.000 & 0.4 &  46  &  14 \\
J2253+1615 & 343.49 & 16.15 & $2.2 \pm 0.1$ &  0.406 & 6.3 & 68 &  20 \\
(3C\,454.3) & & & & & & & \\
\hline
\end{tabular}
\end{center}
\caption{Expected number of neutrino induced muons per km$^2$ per year for 
the sources of the
Second EGRET Catalog together with their right ascension (RA),
declination ($\delta$), 
spectral index ($\gamma$) and fraction of time (${\epsilon}_t$) during which
the source is below the horizon.
The three numbers of neutrino induced muons were found using 
a muon energy threshold of 1 TeV and three different choices of spectral index.}
\label{kn:agn_discr}
\end{table}

\newpage
\clearpage
%
%
\begin{figure}[hhh]
  \begin{center}
    \epsfig{file=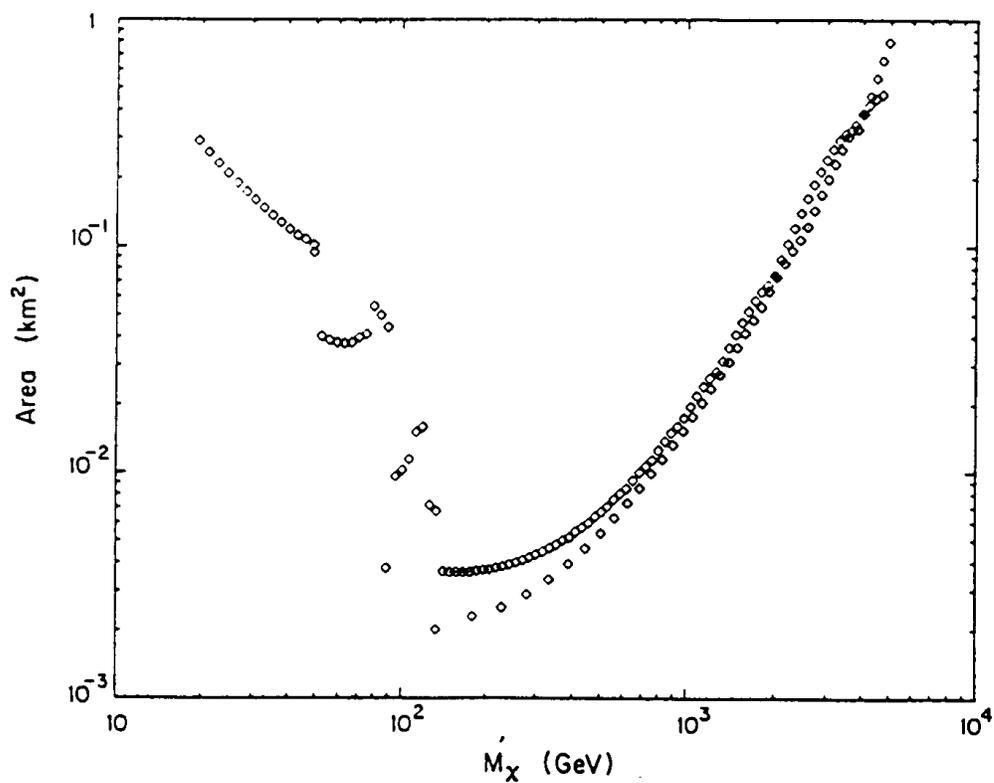,height=0.5\textheight}
    \caption{Detector effective area required to reach a one event per
             year level sensitivity. M$_\chi$ is the neutralino mass.
             More details can be found in \protect\cite{gaisser}}
    \label{fig:wimps}
  \end{center}
\end{figure}
%
%
\begin{figure}[p]
\begin{minipage}[b]{\linewidth}
\centering\epsfig{figure=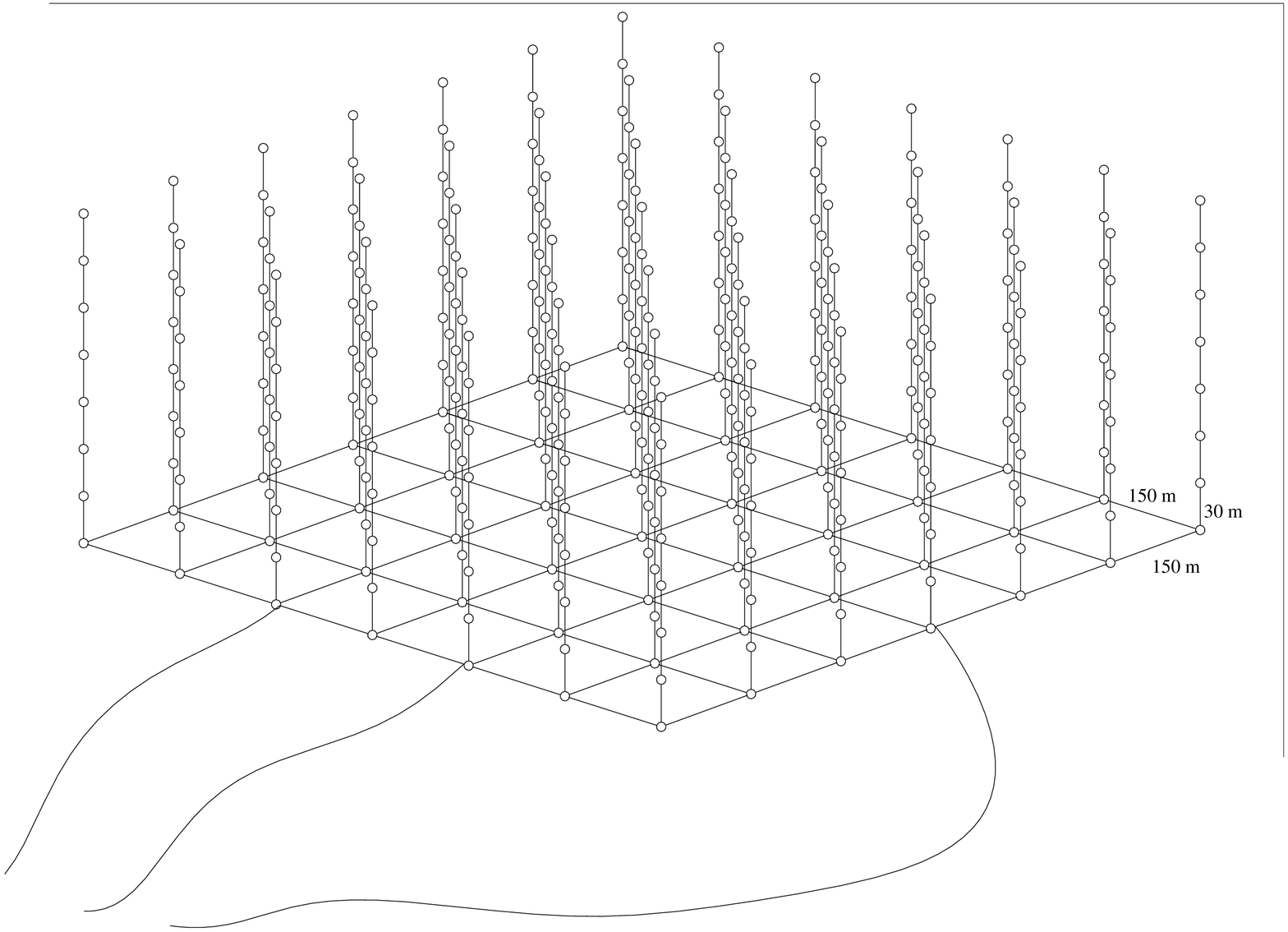,width=\linewidth,angle=270}
\end{minipage} \\
\begin{minipage}[b]{\linewidth}
\centering\epsfig{figure=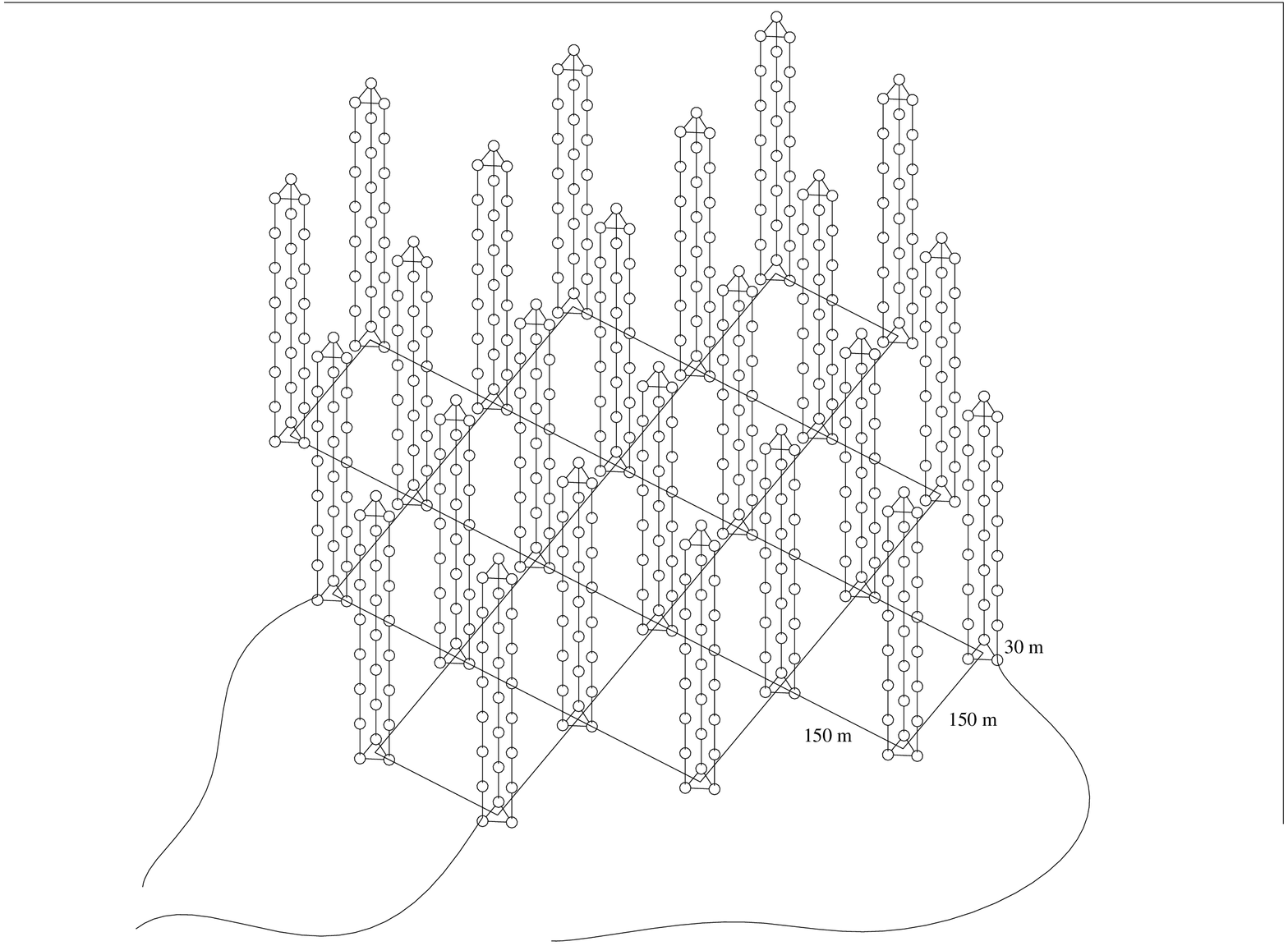,width=\linewidth,angle=270}
\end{minipage} 
\caption{Two possible configuations for a neutrino telescope.}
\label{trames}
\end{figure}     
%
%
\clearpage
\begin{figure}[p]
  \begin{center}
  \mbox{\epsfig{file=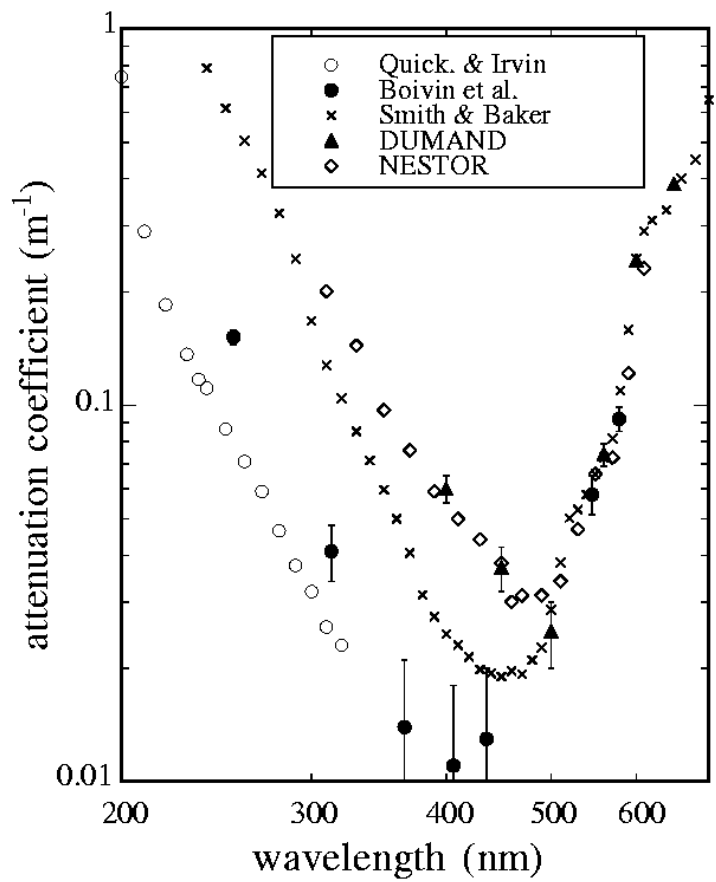,width=10cm}}
  \caption{Attenuation of light in sea water as of function of 
  wave length \protect\cite{price}.}
  \label{attenuation}
  \end{center}
\end{figure}
\begin{figure}[p]
\begin{minipage}[t]{\linewidth}
\centering\epsfig{figure=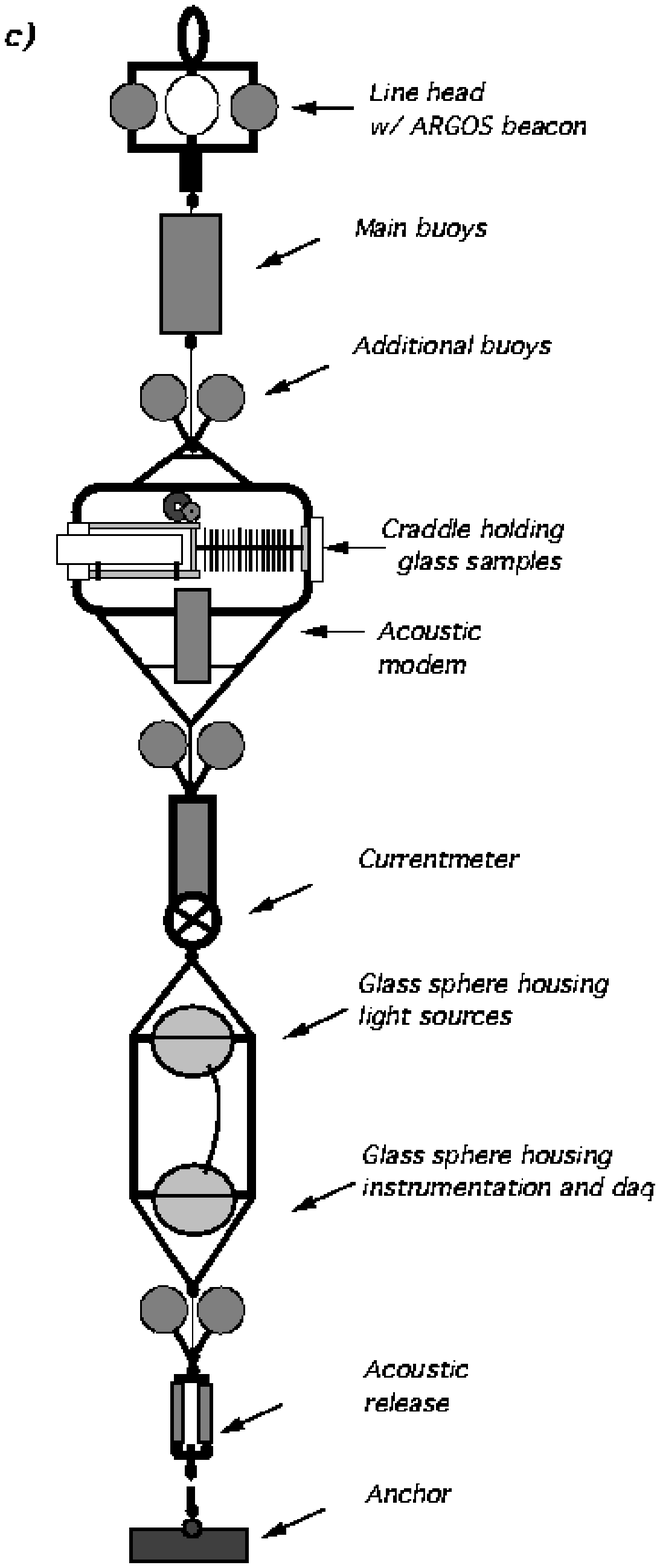,width=.9\linewidth,angle=90}
\end{minipage} \vfill
\begin{minipage}[h]{\linewidth}
\centering\epsfig{figure=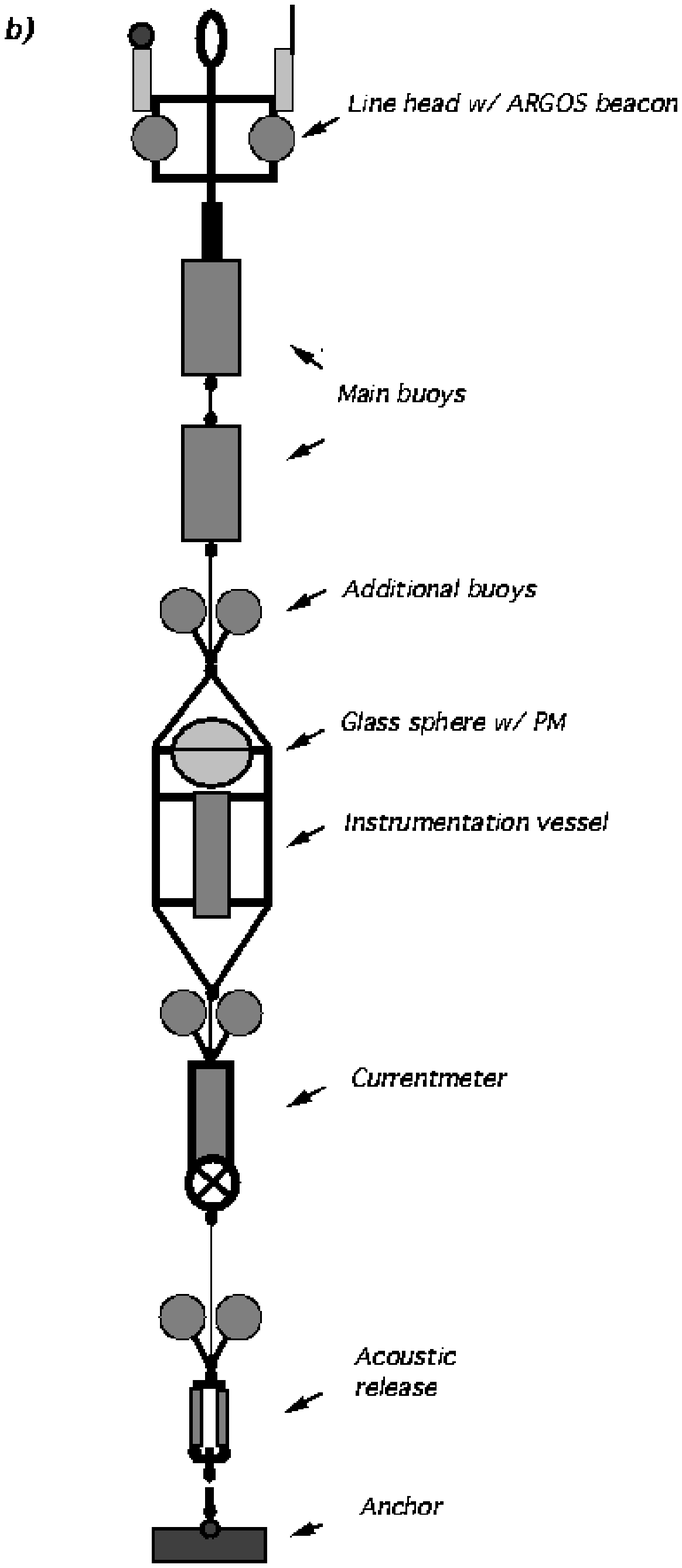,width=.9\linewidth,angle=90}
\end{minipage} \vfill
\begin{minipage}[h]{\linewidth}
\centering\epsfig{figure=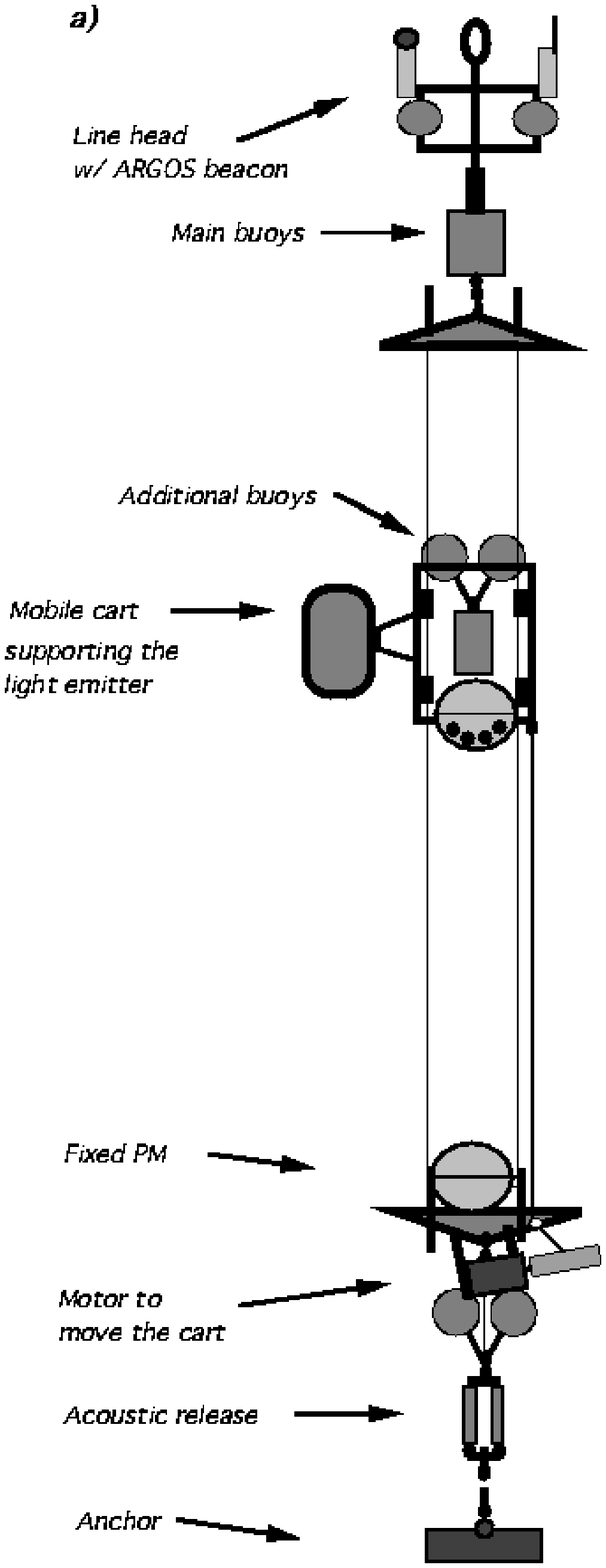,width=.9\linewidth,angle=90}
\end{minipage}
\caption{Schematic mooring lines to measure
   water transparency (a),
   optical background (b),
   biofouling growth (c)} 
\label{fig:tests}
\end{figure}     
\begin{figure}[p]
\begin{center}
(Figure converted to jpeg format)
\caption{Map of the mooring location used for our site study tests
(42$^\circ$50'N-6$^\circ$10'E at a depth of about 2400m).}
\label{carte-toulon}
\end{center}
\end{figure}   
\begin{figure}[p]
\begin{center}
\epsfig{figure=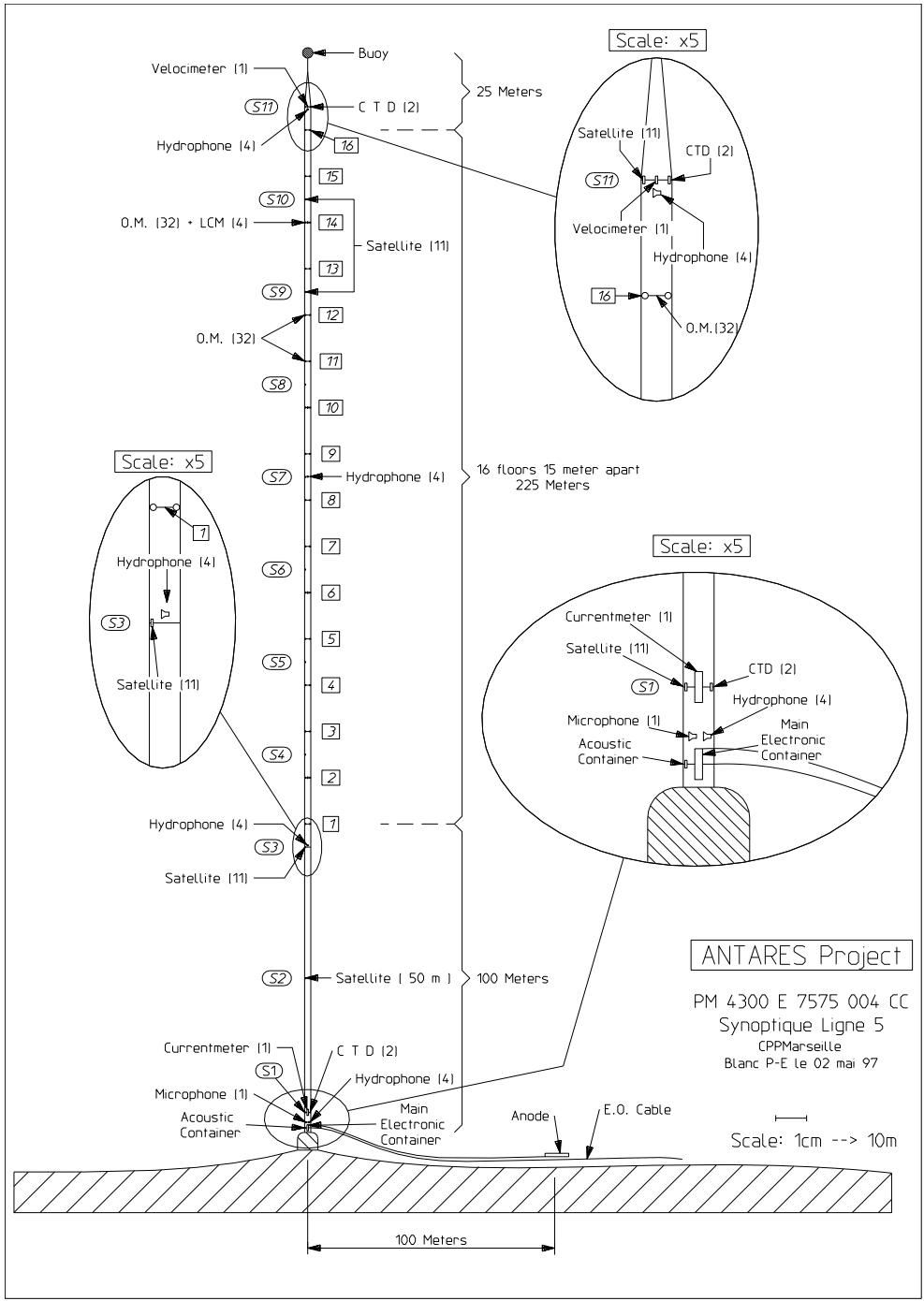,width=0.9\textwidth,angle=180}
\caption{Design of an elementary substructure.}
\label{fig9jjd}
\end{center}
\end{figure}   
\begin{figure}[p]
\begin{center}
(Figure converted to jpeg format)
\caption{Possible set-up of a 3D array of optical modules 
with 3 strings 100 meters apart (not to scale).}
\label{fig10jjd}
\end{center}
\end{figure}   
\begin{figure}[p]
\begin{center}
\mbox{\epsfig{file=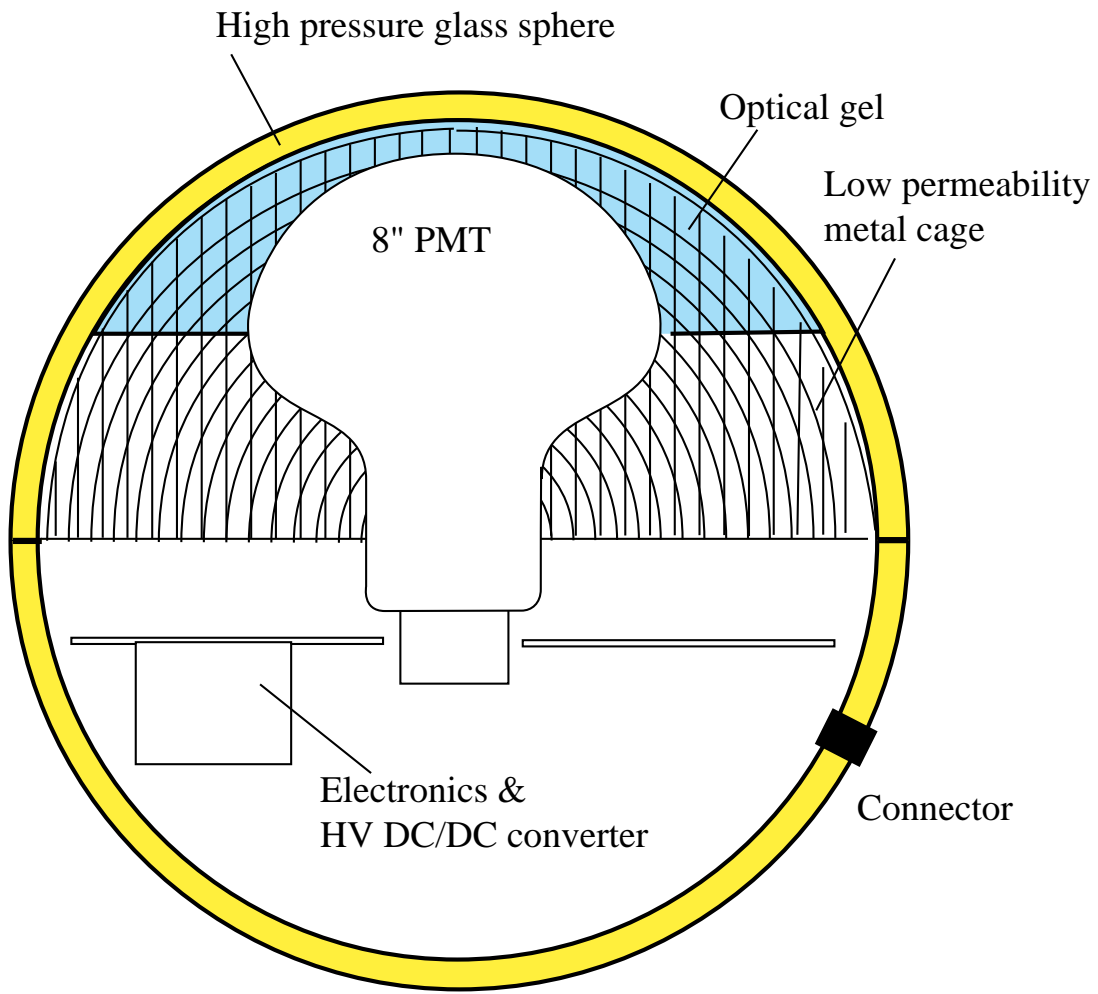,width=15cm}}
\caption{Optical module cross-section.}
\label{modulopt}
\end{center}
\end{figure}   
%
%
%
\clearpage  
\begin{figure}
\begin{center}
  \fbox{
  \begin{tabular}[t]{cc}
    \subfigure[100 GeV muon]
	{\epsfig{figure=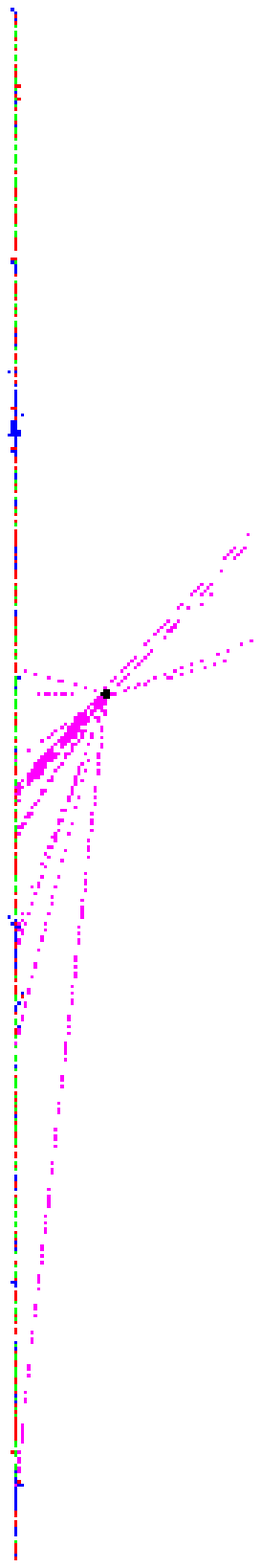,clip=,height=0.5\textheight}}
	&
    \subfigure[10 TeV muon]
	{\epsfig{figure=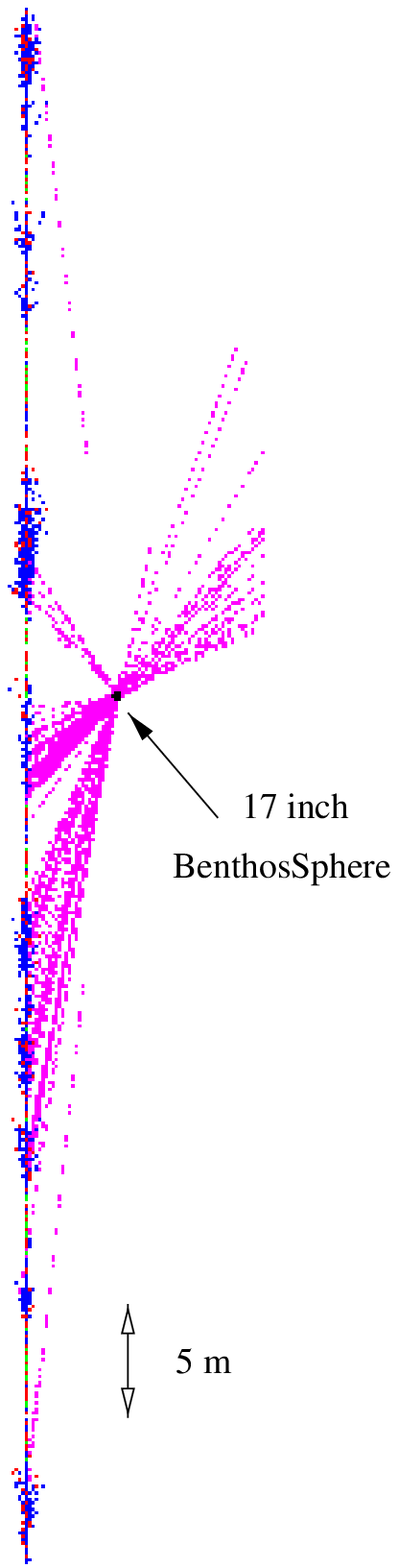,clip=,height=0.5\textheight}} \\
  \end{tabular}}
\end{center}
\caption{Monte Carlo simulation in water of a 100 GeV and 10 TeV muon track
and its induced secondary particles.
We drew only the Cherenkov photons which were able to reach a volume slightly
bigger than the 17 inch Benthos sphere housing a PMT.}
\label{muons}
\end{figure}
\begin{figure}[p]
\begin{center}
\mbox{\epsfig{file=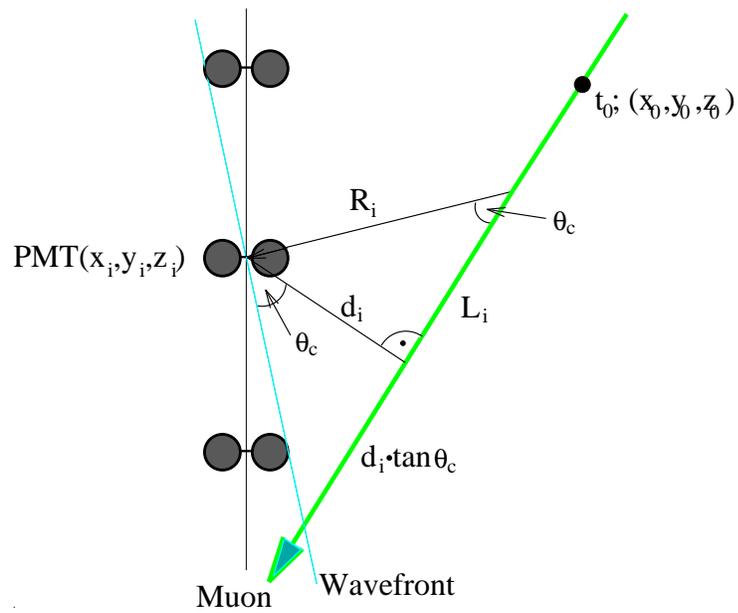,width=10cm}}
\caption{Track reconstruction principle.}
\label{fig:recon}
\end{center}
\end{figure}
\begin{figure}[p]
  \begin{center}
  \mbox{\epsfig{file=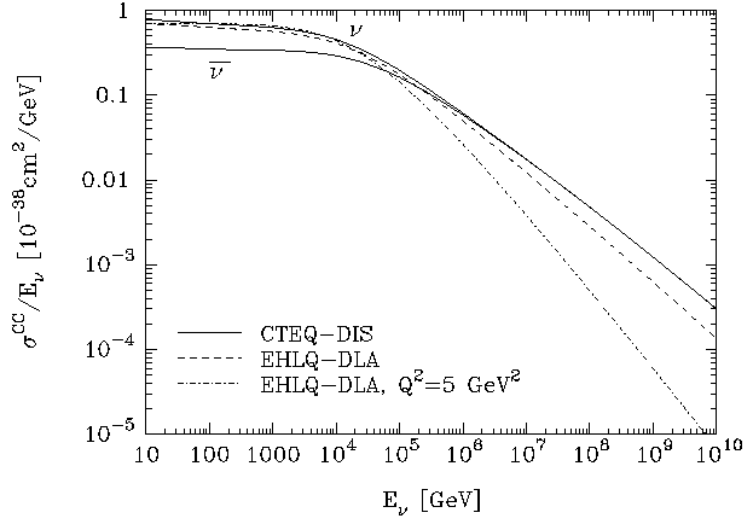,width=10cm}}
  \caption{Energy dependence of the $\nu N$ and $\bar\nu N$ charged current
  cross-sections according to CTEQ3 and EHLQ parton distribution functions
  (taken from \protect{\cite{quigg}}).}
  \label{fig:quigg-4}
  \end{center}
\end{figure}   
\begin{figure}[p]
  \begin{center}
  \mbox{\epsfig{file=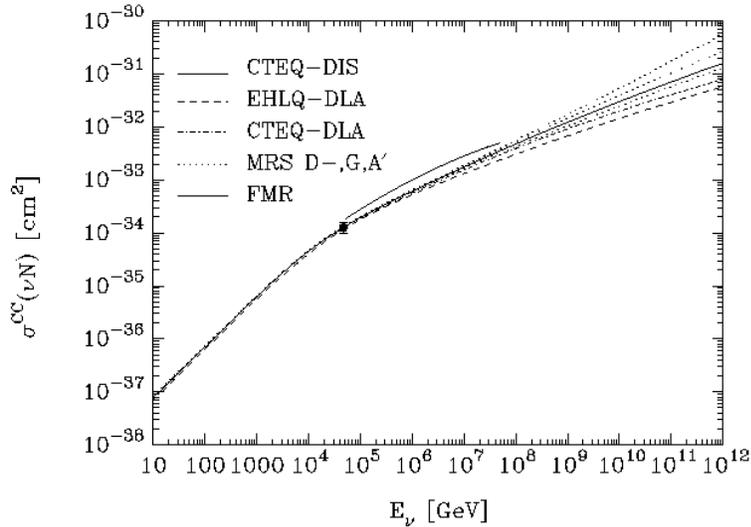,width=10cm}}
  \caption{CC-cross section for $\nu_\mu N$ interactions for different sets of
  parton distribution functions. The data point corresponds to the average of
  measurements by H1 and ZEUS collaborations at HERA
  (taken from \protect{\cite{quigg}}).}
  \label{fig:quigg-5}
  \end{center}
\end{figure}   
\begin{figure}[p]
  \begin{center}
  \mbox{\epsfig{file=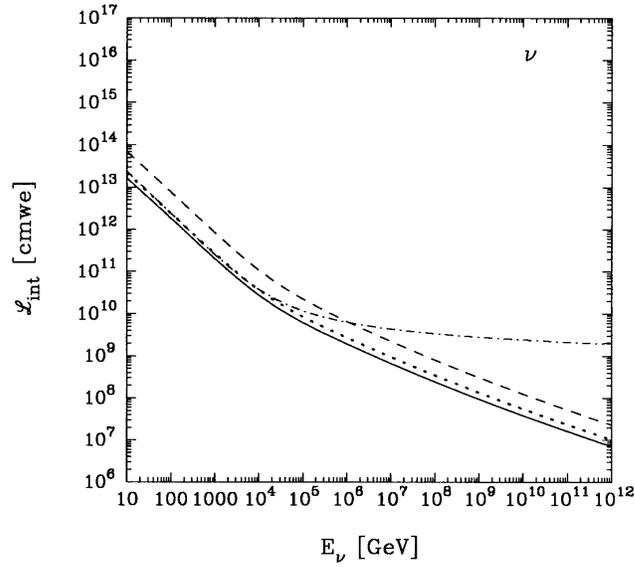,width=9cm}}
  \caption{Interaction lengths for $\nu N$ interactions. When computed with the
  CTEQ3-DIS parton distribution functions: dotted line, CC-interactions; dashed
  line, NC-interactions; solid line CC+NC. The dot-dashed curve corresponds to
  CC-interaction with EHLQ unevolved ($Q^2 = 5$~GeV$^2$) structure functions
  (taken from \protect{\cite{quigg}}).}
  \label{fig:quigg-11}
  \end{center}
\end{figure}   
\begin{figure}[p]
  \begin{center}
  \mbox{\epsfig{file=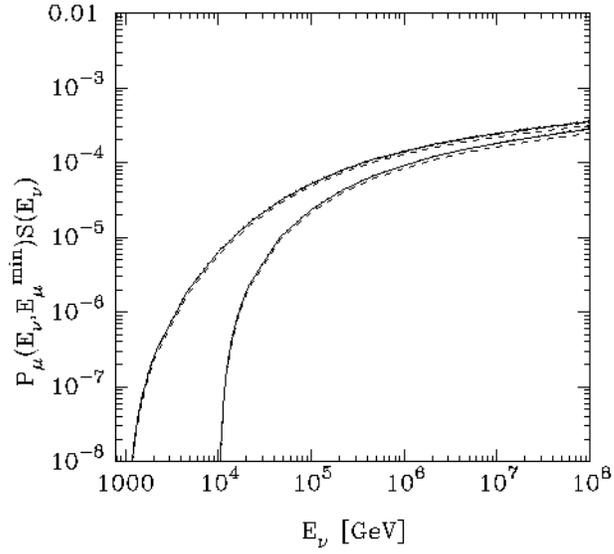,width=9cm}}
  \caption{$P_\mu(E_\nu;E^{min}_\mu)S(E_\nu)$ as a function of $E_\nu$ for
  $E^{min}_\mu$ = 1 TeV and 10 TeV respectively. The curves correspond to
  CTEQ3-DIS (solid) and EHLQ-DLA (dashed). 
  (taken from \protect{\cite{quigg}}).}
  \label{fig:quigg-23}
  \end{center}
\end{figure}   
\end{document}